\documentclass[twocolumn]{aastex631}
\usepackage{bm}

\newcommand\swinburne{Centre for Astrophysics and Supercomputing, Swinburne University of Technology, P.O. Box 218, Hawthorn, Victoria 3122, Australia}
\newcommand{\OzGravMonash}{OzGrav: The Australian Research Council Centre of Excellence for Gravitational Wave Discovery, Clayton VIC 3800, Australia}
\newcommand{\OzGravSwin}{OzGrav: The Australian Research Council Centre of Excellence for Gravitational Wave Discovery, Hawthorn VIC 3122, Australia}

\newcommand\CSIRO{Australia Telescope National Facility, CSIRO, Space and Astronomy, P.O. Box 76, Epping, NSW 1710, Australia}
\newcommand\MQ{Department of Physics and Astronomy and MQ Research Centre in Astronomy, Astrophysics and Astrophotonics, Macquarie University, NSW 2109, Australia}
\newcommand\monash{School of Physics and Astronomy, Monash University, VIC 3800, Australia}

\begin{document}

\title{Search for an isotropic
gravitational-wave background with the Parkes Pulsar Timing Array}

\author[0000-0002-2035-4688]{Daniel J. Reardon}\thanks{E-mail: dreardon@swin.edu.au}
\affiliation{\swinburne}
\affiliation{\OzGravSwin}
\author[0000-0002-9583-2947]{Andrew Zic}\thanks{E-mail: andrew.zic@csiro.au}
\affiliation{\CSIRO}
\affiliation{\MQ}

\author[0000-0002-7285-6348]{Ryan M. Shannon}
\affiliation{\swinburne}
\affiliation{\OzGravSwin}

\author[0000-0003-1502-100X]{George B. Hobbs}
\affiliation{\CSIRO}

\author[0000-0003-3294-3081]{Matthew Bailes}
\affiliation{\swinburne}
\affiliation{\OzGravSwin}

\author[0000-0003-3432-0494]{Valentina Di Marco}
\affiliation{\monash}
\affiliation{\OzGravMonash}

\author[0009-0001-5071-0962]{Agastya Kapur}
\affiliation{\MQ}
\affiliation{\CSIRO}

\author{Axl F. Rogers}
\affiliation{Institute for Radio Astronomy \& Space Research, Auckland University of Technology, Private Bag 92006, Auckland 1142, New Zealand}

\author[0000-0002-4418-3895]{Eric Thrane}
\affiliation{\monash}
\affiliation{\OzGravMonash}

\author[0009-0002-9845-5443]{Jacob Askew}
\affiliation{\swinburne}
\affiliation{\OzGravSwin}

\author[0000-0002-8383-5059]{N. D. Ramesh Bhat}
\affiliation{International Centre for Radio Astronomy Research, Curtin University, Bentley, WA 6102, Australia}

\author[0000-0002-2037-4216]{Andrew Cameron}
\affiliation{\swinburne}
\affiliation{\OzGravSwin}

\author[0000-0002-7031-4828]{Ma\l{}gorzata Cury\l{}o}
\affiliation{Astronomical Observatory, University of Warsaw, Aleje Ujazdowskie 4, 00-478 Warsaw, Poland}

\author[0000-0002-5714-7471]{William A. Coles}
\affiliation{Electrical and Computer Engineering, University of California at San Diego, La Jolla, California, U.S.A.}

\author[0000-0002-9618-2499]{Shi Dai}
\affiliation{School of Science, Western Sydney University, Locked Bag 1797, Penrith South DC, NSW 2751, Australia}

\author[0000-0003-3189-5807]{Boris Goncharov}
\affiliation{Gran Sasso Science Institute (GSSI), I-67100 L'Aquila, Italy}
\affiliation{INFN, Laboratori Nazionali del Gran Sasso, I-67100 Assergi, Italy}

\author[0000-0002-0893-4073]{Matthew Kerr}
\affiliation{Space Science Division, US Naval Research Laboratory, 4555 Overlook Ave SW, Washington DC 20375, USA}

\author[0000-0003-4847-4427]{Atharva Kulkarni}
\affiliation{\swinburne}
\affiliation{\OzGravSwin}

\author{Yuri Levin}
\affiliation{Physics Department and Columbia Astrophysics Laboratory, Columbia University, 538 West 120th Street, New York, NY 10027, USA }
\affiliation{Center for Computational Astrophysics, Flatiron Institute, 162 5th Ave, NY10011, USA}
\affiliation{\monash}

\author[0000-0001-9208-0009]{Marcus E. Lower}
\affiliation{\CSIRO}

\author[0000-0001-9445-5732]{Richard N. Manchester}
\affiliation{\CSIRO}

\author[0000-0001-5131-522X]{Rami Mandow}
\affiliation{\MQ}
\affiliation{\CSIRO}

\author[0000-0002-5455-3474]{Matthew T. Miles}
\affiliation{\swinburne}
\affiliation{\OzGravSwin}

\author[0000-0002-3922-2773]{Rowina S. Nathan}
\affiliation{\monash}
\affiliation{\OzGravMonash}

\author[0000-0003-0289-0732]{Stefan Os{\l}owski}
\affiliation{Manly Astrophysics, 15/41-42 East Esplanade, Manly, NSW 2095, Australia}

\author[0000-0002-1942-7296]{Christopher J. Russell}
\affiliation{CSIRO Scientific Computing, Australian Technology Park, Locked Bag 9013, Alexandria, NSW 1435, Australia}

\author[0000-0002-6730-3298]{Ren\'ee Spiewak}
\affiliation{Jodrell Bank Centre for Astrophysics, Department of Physics and Astronomy, University of Manchester, Manchester M13 9PL, UK}

\author[0000-0001-7049-6468]{Songbo Zhang}
\affiliation{Purple Mountain Observatory, Chinese Academy of Sciences, Nanjing 210008, China}
\affiliation{\CSIRO}

\author[0000-0001-7049-6468]{Xing-Jiang Zhu}
\affiliation{Advanced Institute of Natural Sciences, Beijing Normal University, Zhuhai 519087, China}

\begin{abstract}

Pulsar timing arrays aim to detect nanohertz-frequency gravitational waves (GWs). A background of GWs modulates pulsar arrival times and manifests as a stochastic process, common to all pulsars, with a signature spatial correlation. 
Here we describe a search for an isotropic stochastic gravitational-wave background (GWB) using observations of 30 millisecond pulsars from the third data release of the Parkes Pulsar Timing Array (PPTA), which spans 18 years. 
Using current Bayesian inference techniques we recover and characterize a common-spectrum noise process. Represented as a strain spectrum $h_c = A(f/1 {\rm yr}^{-1})^{\alpha}$, we measure $A=3.1^{+1.3}_{-0.9} \times 10^{-15}$ and $\alpha=-0.45 \pm 0.20$ respectively (median and 68\% credible interval). For a spectral index of $\alpha=-2/3$, corresponding to an isotropic background of GWs radiated by inspiraling supermassive black hole binaries, we recover an amplitude of $A=2.04^{+0.25}_{-0.22} \times 10^{-15}$. However, we demonstrate that the apparent signal strength is time-dependent, as the first half of our data set can be used to place an upper limit on $A$ that is in tension with the inferred common-spectrum amplitude using the complete data set. We search for spatial correlations in the observations by hierarchically analyzing individual pulsar pairs, which also allows for significance validation through randomizing pulsar positions on the sky. For a process with $\alpha=-2/3$, we measure spatial correlations consistent with a GWB, with an estimated false-alarm probability of $p \lesssim 0.02$ (approx. $2\sigma$). The long timing baselines of the PPTA and the access to southern pulsars will continue to play an important role in the International Pulsar Timing Array.
\end{abstract}

\section{Introduction}
\label{sec:intro}

The era of observational gravitational-wave (GW) astronomy commenced with the observation of a stellar-mass binary black hole merger \citep{Abbott+16}. The growing catalog of GW events from ground-based observatories covers relatively high frequencies in a band spanning $\sim 10^1\,$Hz to $\mathcal{O}(10^3)\,$Hz \citep{Abbott+21}. Pulsar timing array (PTA) experiments offer a complementary window into the GW landscape in the nanohertz-frequency band at frequencies $\mathcal{O}(10^{-9})\,$Hz. Sources of GWs at such low frequencies may include supermassive black hole binary (SMBHB) systems \citep[e.g.][]{Rajagopal+95, Wyithe+03, Sesana13, Ravi+15, Burke-Spolaor+19}, primordial quantum fluctuations amplified by inflation \citep[e.g.][]{Grishchuk+05, Lasky+16}, cosmic strings \citep[e.g.][]{Siemens+07, Olmez+10}, or cosmological phase transitions \citep[e.g.][]{Caprini+10, Xue+21}. The highest-amplitude (loudest) signal in this band is expected to be the stochastic background produced by the incoherent superposition of GWs radiated by inspiraling SMBHBs. An isotropic stochastic gravitational-wave background (GWB) is the primary target for GW searches by current PTA collaborations.

A PTA consists of a set of millisecond pulsars (MSPs) that are monitored regularly (cadences of many days to a few weeks), over decade timescales, to take advantage of their long-term rotational stability \citep{Foster+90}. As GWs impart a strain on space-time, they slowly modulate the observed pulse times of arrival (ToAs) from MSPs in an achromatic manner \citep{Sazhin78, Detweiler79}. The ToA modulations induced by an isotropic GWB are predicted to possess a steep power spectral density (PSD) $P \propto f^{-\gamma}$, where $f$ is the GW frequency. This spectrum is derived from the GW strain amplitude spectrum $h_c = A(f/1 {\rm yr}^{-1})^{\alpha}$, where $\alpha = (3 - \gamma)/2$. The characteristic spectral index of an isotropic GWB from inspiraling SMBHBs is $\alpha=-2/3$ and thus $\gamma=13/3$ \citep{Phinney01}. The challenge in detecting this signal is that MSPs are not perfectly stable in their spin and pulse properties and may show intrinsic low-frequency noise that could resemble a GWB for individual pulsars \citep{Shannon+10}. Additionally, plasma in the interstellar medium has a significant effect on radio pulses as it imparts dispersion and scattering delays that depend on the radio wavelength as $\lambda^2$ and $\lambda^{\sim 4}$ respectively \citep[e.g.][]{You+07, Hemberger+08}. Variations in these propagation effects can also produce a red signal in the pulse arrival times \citep[e.g.,][]{Cordes+10,Keith+13}. Other low-frequency noise sources can include errors in the solar system ephemeris (SSE), clock errors, unmodeled variability in the solar wind, or offsets induced by the observing instrumentation \citep{Tiburzi+16}.

The distinct characteristic of an isotropic GWB, which separates it from intrinsic pulsar spin noise and other noise processes, is the spatial correlations in the arrival times. The Hellings--Downs function describes the expected correlation in the pulse arrival times between pairs of pulsars, as a function of their sky separation angle \citep{HellingsDowns}. It is the \textit{fingerprint} of the GWB and is presently the most sought-after signal for a PTA experiment. The sensitivity of a PTA to such spatial correlations depends primarily on the number of pulsar pairs and their sky distribution \citep{Siemens+13, Taylor+16}, which motivates the combination of global PTA datasets under the International Pulsar Timing Array \citep[IPTA;][]{Hobbs+10, IPTA_dr2_gwb} project. Future observations in the high signal-to-noise regime of the GWB may reveal anisotropy resulting from an unresolved local population of SMBHBs \citep{Taylor+13, Mingarelli+17}. Individual supermassive black hole sources may also be observed, with orbital properties inferred through template matching of the GW waveform \citep{Jenet+04, Corbin+10, Ellis13, Zhu+15}.

The Hellings--Downs signature cross correlation is relatively weak, with the average amplitude of the cross correlation being well under 0.25, depending on the distribution of the pulsars on the sky. As the number of pulsars in the array increases, along with the time span and advances in instrumentation that improve the timing precision, the sensitivity of a PTA also increases. We would expect, eventually, to see the same GWB spectrum in the power spectrum for each pulsar, rising above the noise level (which is different for each pulsar). However, the cross correlations may not become evident until the sensitivity of the PTA increases significantly \citep{Pol+21, Romano+21}. Exactly this situation occurred when the North American Nanohertz Observatory for gravitational waves (NANOGrav) detected a ``common-spectrum" process in their pulsars using $12.5\,$yr of MSP timing data, but was unable to establish a significant cross correlation between pulsar pairs \citep{NG12.5yrSGWB}. A similar common-spectrum process was also detected by the other PTAs without obvious cross correlations \citep{PPTA_dr2_crn, EPTA_dr2_crn}. The IPTA, using a union of earlier data releases from each of these PTAs, also identified a common-spectrum process, notably with higher significance than the individual PTA constituents of the dataset \citep{IPTA_dr2_gwb}.

 The origin of the common-spectrum process is unclear, in part because these recent results were in apparent tension with the 95\% confidence upper limits placed by some PTA analyses using earlier data releases \citep[e.g.][]{Shannon+15, NG11yrSGWB}. Unmodeled SSE errors \citep[][]{Bayesephem}, choices of noise models and priors \citep{Hazboun+20b}, and finite numbers of pulsars \citep{Johnson+22} have been proposed to have been factors in these earlier upper-limit results. The NANOGrav analysis, and earlier works \citep[][]{NG11yrSGWB, Bayesephem}, demonstrated that SSE errors could be contributing to the measured common noise, and consequently, techniques were developed to account for such errors.  The PPTA analysis found that the ``common process'' model does not discriminate between common and spectrally similar noise processes \citep{Zic+22}, which was addressed in \citet{Goncharov+22}. Depending on the noise modeling framework, a recovered common-spectrum process could contain various non-GW sources of noise. For example, the common spectrum could include pulsar timing noise and/or errors in the correction of interstellar medium effects (dispersion and scattering), which would be uncorrelated between pulsars and could have a similar power-law red spectrum. It could also include SSE errors, which would be correlated between pulsars and could have a wide range of spectral distributions \citep[][]{Guo+19,Bayesephem}. Evidently, the origin of the common spectrum will not be resolved until a significant cross correlation is obtained.


For this work we have used the third data release of the PPTA \citep{PPTA-DR3_data} to search for and characterize a common-spectrum stochastic process and spatial correlations, with particular focus on the GWB. In a coordinated effort, the NANOGrav, EPTA and Indian Pulsar Timing Array \citep{InPTA}, and Chinese Pulsar Timing Array \citep{CPTA} collaborations have also recently searched their respective data sets for a GWB \citep{ng_15yr_gwb, epta_inpta_gwb, cpta_gwb}. In Section \ref{sec:methods}, we describe the data analysis methods and the models we consider. The results are presented in Section \ref{sec:results} and the implications are discussed further in Section \ref{sec:discussion}. We conclude in Section \ref{sec:conclusions}.
\\
\\
\section{Methods}
\label{sec:methods}
The analysis presented here is performed on the third PPTA data release (PPTA-DR3) and individual pulsar noise analyses. The data release adds $\sim 3$ yr of timing baseline compared to PPTA-DR2 \cite[][]{ppta_dr2} and includes observations taken with the new ultrawide-band low receiver \citep[UWL;][]{uwl}, which offers wider bandwidths than previous observations. This is useful for measuring interstellar medium effects and obtaining arrival times with higher precision. The data release \citep{PPTA-DR3_data} and noise analyses \citep{PPTA-DR3_noise} are presented in companion papers.  While PPTA-DR3 contains observations from 32 pulsars, in this work we only consider 30 of them. The globular cluster MSP PSR~J1824$-$2452A was excluded as it contains steep-spectrum red noise that is too strong for this pulsar to be sensitive to a common process. The noise is likely intrinsic to the pulsar \cite[][]{Shannon+10}, although globular-cluster dynamics may contribute as well. We also exclude PSR~J1741$+$1351 from the analysis. It was only added to the PPTA as the UWL was commissioned, and observed with low priority.  In the current data release, only $16$ observations were available, resulting in a data set that would not be sensitive to any GWB signal. These observations could, however, become useful as part of future IPTA data sets.

\subsection{Inference and PTA likelihood}

We use Bayesian inference\footnote{For an overview of the principles of Bayesian inference in gravitational-wave astronomy, see~\cite{Thrane+19} and~\cite{Taylor21}.} to search for and characterize noise and signals in our data set, following previous PTA analyses \citep{NG12.5yrSGWB, PPTA_dr2_crn, EPTA_dr2_crn, IPTA_dr2_gwb}.
PTAs observe offsets between the measured pulse arrival times  $\bm{\delta t}$ and the arrival times predicted by the models.

The time-domain likelihood of the observed residuals given a model is described by a Gaussian likelihood function, $\mathcal{L}(\bm{\delta t} | \bm{\theta})$, where $\bm{\theta}$ is the vector of model parameters \citep{vanHaasteren+09}. 
The likelihood is multivariate with respect to the number of observations of timing residuals. 
The implementation of the likelihood is described in \citet{NG9yr} and \citet{Taylor+17}, and it is given by
\begin{equation}
    \mathcal{L}(\bm{\delta t} | \bm{\theta}) = \frac{\exp\left(  -\frac{1}{2} (\bm{\delta t} - \bm{\mu})^T \bm{C}^{-1}(\bm{\delta t}-\bm{\mu}) \right)}{\sqrt{\det\left( 2\pi \bm{C}\right)}},
\end{equation}    
where $\bm{\mu}(\bm{\theta})$ represents the model prediction for timing residuals, and the covariance matrix is $\bm{C}(\bm{\theta})$.  The diagonal elements in the covariance matrix represent the temporally uncorrelated noise, which is referred to as white.
Temporally correlated stochastic processes, referred to as red, could have been represented by off-diagonal elements in $\bm{C}$.
However, to avoid a computationally expensive matrix inversion, temporally correlated stochastic processes are modeled as Gaussian processes \citep{Lentati+13, vanHaasteren+14} in $\bm{\mu}$:
\begin{equation}\label{eqn:single_psr_likelihood}
    \bm{\mu} = \bm{F}\bm{a} + \bm{M}\bm{\epsilon}.
\end{equation}
where the design matrix $\bm{M}$ and the corresponding coefficients $\bm{\epsilon}$ are the timing model contributions, and the red processes are modeled using the Fourier sine and cosine basis functions in $\bm{F}$ and amplitudes $\bm{a}$.
We emphasize that $(\bm{a}$ and $\bm{\epsilon})$ are a subset of the parameters in the model $\bm{\theta}$.
They are nuisance parameters and can be analytically marginalized over a Gaussian prior $\pi(\bm{a},\bm{\epsilon}|\bm{\theta^{'}})$.   
It is common to assume that there is a relationship between the amplitudes of the Fourier components. 
In this case, $\bm{\theta^{'}}$ are hyperparameters that govern the spectra of temporal correlations, such as the power-law amplitude and the slope of $P(f)$ of pulsar-intrinsic red noise, or the gravitational-wave background.
The prior on $\bm{a}$ joins single-pulsar likelihoods into a joint posterior distribution
\begin{equation}
    \mathcal{P}(\bm{\theta}, \bm{\theta^{'}} | \bm{\delta t}) \propto \mathcal{L}(\bm{\delta t} | \bm{\theta}) \pi(\bm{a},\bm{\epsilon}|\bm{\theta^{'}}) \pi(\bm{\theta^{'}}).
\end{equation}
The covariance matrix of $\pi(\bm{a},\bm{\epsilon}|\bm{\theta^{'}})$ has elements
\begin{equation}
\label{eqn:covariance}
    \pi_{(ai),(bj)} = P_{ai}\delta_{ab}\delta_{ij} + \Gamma_{ab} P_{i}\delta_{ij},
\end{equation}
where $a$ and $b$ are pulsar indices, and $i$ and $j$ are frequency indices. $P_{ai}$ is the PSD of the noise intrinsic to pulsar $a$ at a frequency bin $i$, $P_i$ is the PSD of a spatially correlated signal at the frequency $i$, and $\Gamma_{ab}$ is the overlap reduction function that determines the degree of this correlation between pulsars $a$ and $b$.

In case of an isotropic GWB from circular SMBHBs, $\Gamma_{ab}$, also referred to as the overlap reduction function, is given by the \cite{HellingsDowns} curve:
\begin{equation}
\label{eqn:hd}
    \Gamma_{ab} = \frac{1}{2}\delta_{ab} + \frac{1}{2} - \frac{x_{ab}}{4} + \frac{3}{2}x_{ab}\ln x_{ab},
\end{equation}
where $x_{ab} = (1 - \cos\zeta_{ab})/2$, $\delta_{ab}$ is the Kronecker delta function, and $\zeta_{ab}$ is the sky separation angle for a given pair of pulsars.

We evaluate the posterior probability $\mathcal{P}(\bm{\theta}, \bm{\theta^{'}} | \bm{\delta t})$ using the \textsc{enterprise} package \citep{Enterprise} and a parallel-tempered Markov Chain Monte Carlo (MCMC) algorithm \citep[\textsc{ptmcmcsampler};][]{PTMCMC}. For more details about modeling Gaussian processes in PTAs, see \citet{Lentati+13} and  \citet{vanHaasteren+14}.  

Previous GWB analyses have assessed the properties of the dominant temporally correlated and spatially uncorrelated \textit{common-spectrum} process. This process was considered as the null hypothesis in searches for the spatially correlated GWB signal. The PSD of temporal correlations can be modeled as a power law parameterized by an amplitude ($A$) and spectral index: 
\begin{equation}
\label{eqn:crn}
P(f) = \frac{ A^2 }{12 \pi ^2} \left(\frac{f}{f_{\rm yr}} \right)^{-\gamma},
\end{equation}
where $\gamma>0$. The models of common noise in this form are added to the PTA likelihood and analyzed simultaneously with all red-noise processes identified in single pulsars. The support for the proposed common-noise processes is quantified primarily through the Bayesian odds ratio (equivalent to the Bayes factor $\mathcal{B}$ for equal prior odds), which can be estimated using the product-space sampling method \cite[][]{NG9yr, Carlin+18}

The number of Fourier frequency components used to model the common power-law processes can be set by the sensitivity of our PTA, which is white-noise dominated near $f \sim 1/(240\,{\rm days})$ \citep[as determined based on a broken power-law analysis, e.g.][]{NG12.5yrSGWB}. Accordingly we set the number of components as $N_{\rm comp} = \lfloor T_{\rm span}/(240\, {\rm days}) \rfloor = 28$, where $T_{\rm span}=6605\,{\rm days}$ is the total observing span of PPTA-DR3. We note that the use of a broken power-law model to set the number of significant Fourier components \citep{NG12.5yrSGWB} could highlight excess unmodeled (non-power-law) noise at higher frequencies. If the noise is correctly characterized, and the signal is a power law, then the data will not favor a turnover frequency and should instead become insensitive to a turnover for frequencies that are white-noise dominated.

Under the model of a fixed spectral index $\gamma = 13/3$ \cite[the spectral index expected for a GWB produced by circular SMBHBs, with their inspirals driven by GW radiation at the frequencies of interest; ][]{Phinney01} and zero spatial correlation, the PTA likelihood can be factorized into a product of likelihoods from individual pulsars \citep[e.g.][]{Taylor+22}. In this way, a $\gamma = 13/3$ process can be introduced to the noise models of individual pulsars and the resulting posterior probability density functions (PDFs) for the process amplitude can be multiplied to produce a posterior for the whole array. We use this factorized-likelihood approach to efficiently assess the preferred amplitude of a $\gamma = 13/3$ process in our data, and to estimate the model support under an assumed prior range using the Savage--Dickey method for computing the Bayes factor $\mathcal{B}$ \citep{Dickey71}.

Hereafter, when referring to specific models, we label the amplitude of an uncorrelated common-red-noise (CRN) process as $A^{\rm CRN}$, and when the spectral index is fixed at $\gamma = 13/3$, we write $A^{\rm CRN}_{13/3}$. For a $\gamma = 13/3$ process in a single pulsar (i.e. not common to the PTA), we use $A_{13/3}$. A process with Hellings--Downs correlations included in the covariance matrix is denoted $A^{\rm HD}$ with a varied spectral index and $A^{\rm GWB}$ when the index is $\gamma = 13/3$.

\subsubsection{Defining prior ranges}

The choice of prior has a dramatic effect on the Bayesian evidence of common-noise processes, and therefore Bayes factors used for model comparison \citep[][]{Zic+22}. It is important for priors to encompass the likelihood support, but prior ranges that are too broad have been shown to induce spurious support for certain models. This problem affects the inference of common uncorrelated noise processes in PTAs, and can lead to very high Bayesian support for a common process where none exists \citep{Zic+22}. To mitigate the possibility of spurious support, we trialed different priors in our analysis. For the factorized-likelihood analysis we adopt a log-uniform prior on the amplitude in the range $-20 \leq \log_{10} A_{13/3} \leq -11$ to be consistent with earlier analyses \citep{NG12.5yrSGWB, PPTA_dr2_crn}. However, our data are not sensitive to signals below $\log_{10} A_{13/3} \lesssim -16.5$ even for very steep spectral indices ($\gamma \sim 7$). We therefore adopt $\log_{10} A_{13/3} \geq -18$ as a conservative lower bound for the remainder of our analyses. The prior for the common-noise spectral index is uniform in $ 0 \leq \gamma \leq 7$.

To set the prior ranges for the individual pulsar noise parameters, we take the central $99.7\%$ credible interval from our single-pulsar noise analysis posteriors \citep{PPTA-DR3_noise}, and add a generous buffer region such that pulsar noise properties are allowed to vary substantially under a common-noise model. These ranges are $\log_{10} A_{0.15} - 2 \leq \log_{10} A \leq \log_{10} A_{99.85} + 1$ and $\gamma_{0.15} - 0.5 \leq \gamma \leq \gamma_{99.85} + 0.5$, where the parameter subscripts denote the percentile of the posterior distribution for this parameter. This ensures that we do not force any pulsars to have intrinsic noise if it is instead attributed to a common process and the prior range encompasses all likelihood support from the data. PSR~J1643$-$1224 is the only pulsar with a lower bound $\log_{10} A \geq -17$, with a red noise prior constructed in this way of $-15.2 < \log_{10} A < -11.5$ as it is observed to have shallow-spectrum noise ($\gamma \leq 2.3$ with 99.7\% confidence). By restricting the prior ranges in this way, we reduce the risk of spurious measurements \citep{Zic+22} of common noise and are able to accelerate the inference. We are confident in the recovered parameters using this methodology as the shapes and values of the inferred marginal posterior PDFs using these prior ranges are consistent with those recovered using standard priors.

\subsection{Noise and Signal Models} \label{sec:models}

\subsubsection{Common processes}

The term \textit{common noise} refers to any noise process (or potentially unmodeled deterministic signal), correlated or uncorrelated, that is characterized by the same power spectral density for all pulsars. However, common noise could be recovered from similar power spectral densities in most pulsars, and as such, examples include not only GWs but also potentially unmodeled instrumental offsets, the solar wind; or errors in the clock correction, Earth orientation parameters, or SSE. If intrinsic pulsar timing noise (rotational irregularities) is generated by the same physical mechanism in multiple pulsars, the spectral properties of this random process could be recovered via inference of a common-spectrum noise process but should be spatially uncorrelated, unlike the GWB. The variety of potential common-noise sources highlights the need to measure the signature spatial correlations in order to detect a GWB. The power spectrum of a power-law common uncorrelated process is given by Equation \ref{eqn:crn}.

\subsubsection{Searching for spatial correlations}
\label{sec:corr_method}

We conduct two distinct classes of searches for the Hellings--Downs correlations inherent to a GWB. 
First, we replicate the methods of some previous searches \citep[e.g.][]{NG12.5yrSGWB, PPTA_dr2_crn,EPTA_dr2_crn, IPTA_dr2_gwb}, by constructing the global PTA likelihood for all pulsars and introducing a Hellings--Downs correlated noise term to the likelihood (see Equation \ref{eqn:covariance}).  For this model, both the autocorrelations and cross correlations are considered and inform the parameter inference. The sensitivity of the PPTA data set is predicted to be dominated by the autocorrelations \citep{Spiewak+22}, so we could expect the spectral properties of such a model to closely resemble the uncorrelated common process. While pulsar white-noise terms are usually fixed at their maximum likelihood values (as they are less covariant with the GWB), we numerically marginalized over all red-noise terms for all pulsars in the array, as these noise terms could be highly covariant with any common signal. This results in a high-dimension parameter space that needs to be explored. In the case of PPTA-DR3 the parameter space spans 260 dimensions.

We also present a new search method that increases the efficiency of the inference with a hierarchical inference approach. This is important because with our large data volume ($\sim 1.2 \times 10^5$ ToAs) and parameter space ($\sim 260$ free parameters) it takes of order CPU years to collect a sufficient number of independent MCMC samples for standard full-PTA correlation analyses, which makes it costly to determine false-alarm probabilities with bootstrap methods. We first search for common-spectrum noise process in pulsar pairs individually. We measure the $\log_{10}A_{13/3}$ and correlation coefficient, $\Gamma$, for each of the $435$ unique pulsar pairs ($N_{\rm pair} = N_{\rm psr}(N_{\rm psr} - 1)/2$, for $N_{\rm psr}$ pulsars) in our data set. From the posterior samples, we derive a smooth PDF, $p(\log_{10}A_{13/3}, \Gamma)$, using a two-dimensional Gaussian kernel density estimate (KDE). We account for the bounded domains $-1 \leq \Gamma \leq 1$, and $-18 \leq \log_{10}A \leq -14$ by mirroring the samples at these boundaries to partially overcome bias in the density estimator introduced by the boundary discontinuity \citep{Schuster+85}. 

The standard deviation of the Gaussian kernel (for all pairs) was $\sigma_{\rm KDE}(\log_{10}A) = 0.05$ for amplitude (the standard deviation of our measured common process) and $\sigma_{\rm KDE}(\Gamma) = 0.27$ for correlation coefficient (the minimum observed standard deviation for an individual pair in our sample). We also tried using a different $\sigma_{\rm KDE}(\Gamma)$ kernel width for each pair, computed using Silverman's rule \citep[because we observe that the distributions are unimodal;][]{Silverman86}, and we observe consistent results.

From these resulting PDFs, we reweight the posterior samples \cite[using a form of importance sampling;][]{Gelman+95, Goncharov+22, Hourihane+22, Taylor+22} to recover the probability density for $\Gamma$ that corresponds to the posterior probability density of  $\log_{10}A^{\rm CRN}_{13/3}$ that we measured from our factorized-likelihood analysis. After this reweighting, the $435$ PDFs for $\Gamma$, denoted $p(\Gamma)$, assume a common distribution on $\log_{10}A$. These PDFs can then be analyzed to assess the model support for correlations such as the Hellings--Downs ORF expected of the isotropic GWB. The measurements will not be completely independent, and we address the implications of this below.

We also measure $p(\Gamma)$ for the amplitude fixed at our maximum likelihood $\log_{10}A^{\rm CRN}_{13/3}$, which is equivalent to the reweighted measurements if the correlations do not evolve across the uncertainty of the recovered common noise. These fixed-amplitude measurements are more computationally efficient, particularly for pulsars with low likelihood support at the amplitude of interest. We therefore obtain these fixed-amplitude correlation measurements from independent MCMC chains so that they can also serve as a consistency test, sensitive to errors in the KDE due to a finite number of samples. 

We quantify the support for a Hellings--Downs ORF over uncorrelated common noise by computing the likelihood of each model. The Hellings--Downs likelihood is defined to be the product of the probabilities of observing $\Gamma = \Gamma_{ab}$ for each pair
\begin{equation}
    \mathcal{L}^{\rm HD} = \prod_{k=1}^{N_{\rm pair}} p_k(\Gamma = \Gamma_{ab})
\end{equation}
where the $k^{\rm th}$ pair involves pulsars $a$ and $b$, $N_{\rm pair}=435$, and $\Gamma_{ab}$ is given by Equation \ref{eqn:hd}. Similarly, the likelihood of zero correlations for the observed common noise is
\begin{equation}
    \mathcal{L}^{\rm CRN} = \prod_{k=1}^{N_{\rm pair}} p_k(\Gamma = 0).
\end{equation}
The support for Hellings--Downs over zero correlations can be quantified using the likelihood ratio $\Delta \mathcal{L}^{\rm HD}_{\rm CRN} = \mathcal{L}^{\rm HD} / \mathcal{L}^{\rm CRN}$. This could be interpreted as an estimated Bayes Factor $\mathcal{B}^{\rm HD}_{\rm CRN}$ because the ORFs are effectively zero-parameter models. However, because of the weaknesses in this method (e.g. assuming the measured correlations are independent, and KDE biases), the \textit{ad-hoc} estimator $\Delta \mathcal{L}^{\rm HD}_{\rm CRN}$ is not as well-defined as a Bayes factor, and we instead interpret its statistical significance by generating noise realizations from the data itself. 

The method allows for the rapid computation of false-alarm probabilities by recalculating the likelihood statistics after assuming random positions for the pulsars on the sky. This process is referred to as sky scrambling \cite[][]{Cornish+16, Taylor+17}. The number of possible independent skies (leading to orthogonal sets of samples from the Hellings--Downs ORF) is limited to $N_{\rm sky} = N_{\rm psr}(N_{\rm psr} - 1)$, for $N_{\rm psr}$ pulsars with equal sensitivity. However, we are able to generate a larger number of skies by relaxing the strict requirement for independence. We instead consider randomized skies where the normalized inner product of the ORF samples from any two skies is $\leq 0.2$ \cite[e.g.][]{Cornish+16, Taylor+17}.

Our hierarchical inference scheme can also be simply extended to search for anisotropy in the GWB \citep{Taylor+13}, or to search for a signal with a completely different spatial correlation signature. It could also be used to estimate the amplitude of a correlated signal independently from the autocorrelations (provided a reliable reconstruction of the full $( \log_{10}A, \Gamma )$ parameter space can be made for each pair).

Pulsars have different sensitivities to common-noise processes due to variations in their timing precision and intrinsic red-noise properties. As a result, the effective number of pulsar pairs in the array is less than the total. We can also use the PDFs for $\Gamma$ to compute the number of effective pulsar pairs $n_{\rm eff}$ using
\begin{equation}
\label{eqn:neff}
    n_{\rm eff} = \frac{\left( \sum^{N_{\rm pair}}_{k=1}\sigma_k^{-2} \right)^2}{\sum^{N_{\rm pair}}_{k=1}\sigma_k^{-4} },
\end{equation}
where $\sigma_k$ is the uncertainty in the angular correlation for the $k^{\rm th}$ pulsar pair. We define $\sigma_k$ as the half-width of the 68\% credibility interval of the correlation coefficient posterior for pulsar pair $k$. This is more conservative than using the standard deviation of the posterior samples as $\sigma_k$ because the distributions are often asymmetric. 

\section{Results}
\label{sec:results}

\subsection{Factorized-likelihood analysis}

\begin{figure*}
\centerline{\includegraphics[width=0.7\textwidth]{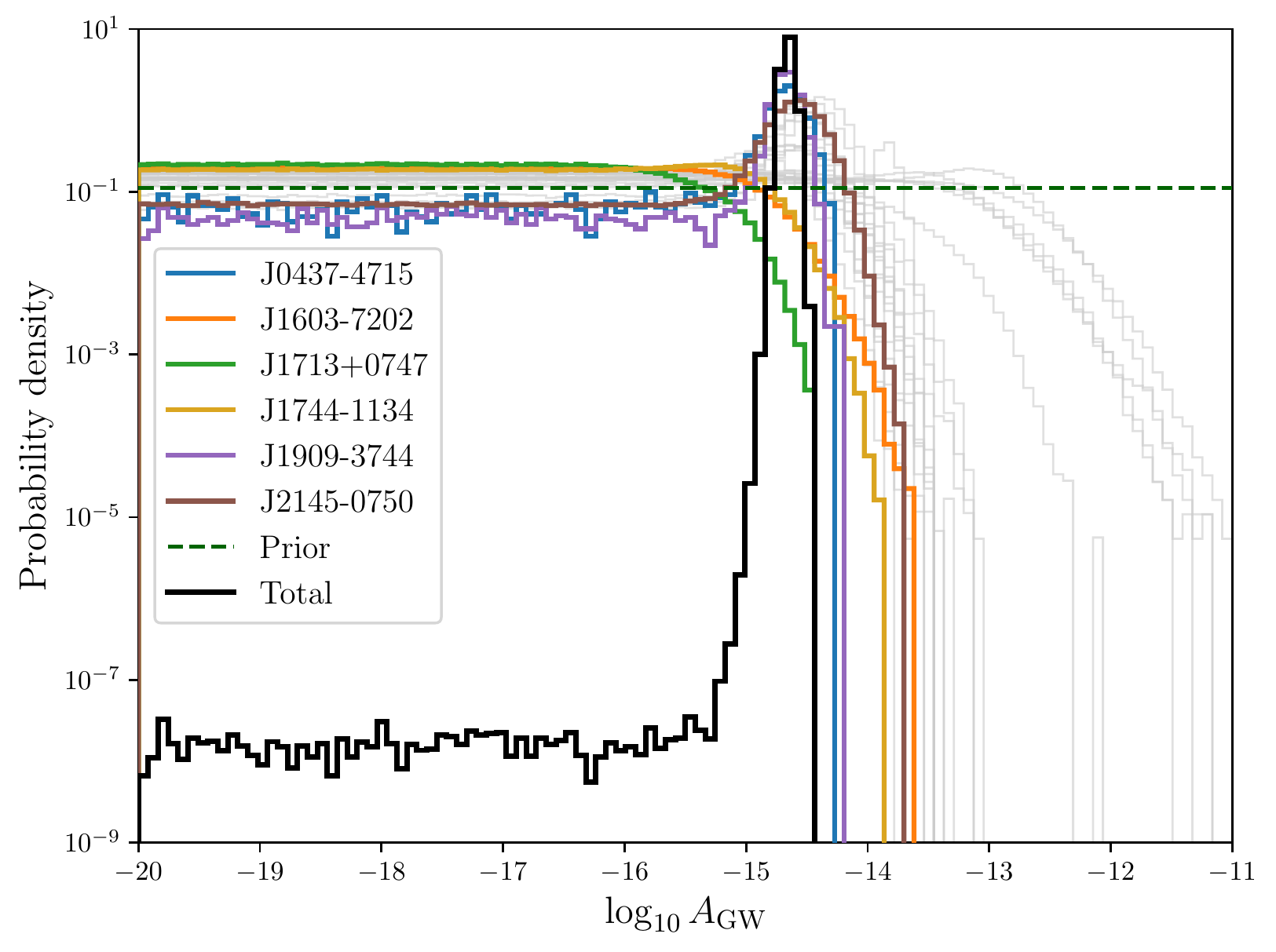}}
\caption{Factorized-likelihood analysis on the $\log_{10} A_{13/3}$ from the PPTA-DR3 pulsars. We highlight the three pulsars showing the highest (PSRs~J1909$-$3744, J0437$-$4715, and J2145$-$0750) and lowest (PSRs~J1744$-$1134, J1603$-$7202, and J1713$+$0747) consistency with the inferred common process with colored histograms. Other pulsars are presented as light-gray histograms. The prior density is shown with the green dashed line, and the total factorized-likelihood constraint for $\log_{10} A^{\mathrm{CRN}}_{13/3}$ (the product of all other histograms) is shown by the black line.}
\label{fig:factorised_likelihood}
\end{figure*} 

We first searched for a purported common red process using the factorized-likelihood technique \citep{Taylor+22}. For each pulsar, we include an achromatic process with a fixed spectral index of $\gamma=13/3$, while the  (log) amplitude for this process, with a prior in the range $-20 \leq \log_{10}A_{13/3} \leq -11$, was sampled simultaneously with the full single-pulsar noise models. The posterior probability density for each pulsar (light-gray and colored lines) is shown in Figure \ref{fig:factorised_likelihood}. The probability density for the log amplitude of a \textit{common process} is found by taking the product of these individual pulsar posteriors. In this way, the log amplitude can be interpreted as a probability-weighted mean. The resulting distribution gives $\log_{10}A^{\mathrm{CRN}}_{13/3} = -14.69 \pm 0.05$, where, here and throughout, the uncertainties represent the central 68\% credibility interval unless otherwise stated.

The PPTA data as a whole support the inclusion of a $\gamma = 13/3$ process at this amplitude, with a strong Savage--Dickey Bayes factor, $\log_{10}\mathcal{B} = 6.8$. By reducing the prior range to our nominal $-18 \leq \log_{10} A_{13/3} \leq -11$, we find that the evidence decreases to $\log_{10}\mathcal{B} = 4.8$, indicating that this statistic alone is a poor metric for determining if a common-noise process is present in the array. We stress that, while the evidence for a common-noise process varies under different assumed prior ranges, the recovered characteristics do not, because the likelihood support under a common-noise model is encompassed for all pulsars for all prior ranges considered.

\subsection{Noise consistency metrics}

\begin{figure*}
\centerline{\includegraphics[width=0.95\textwidth, trim= 0 0 0 0 , clip]{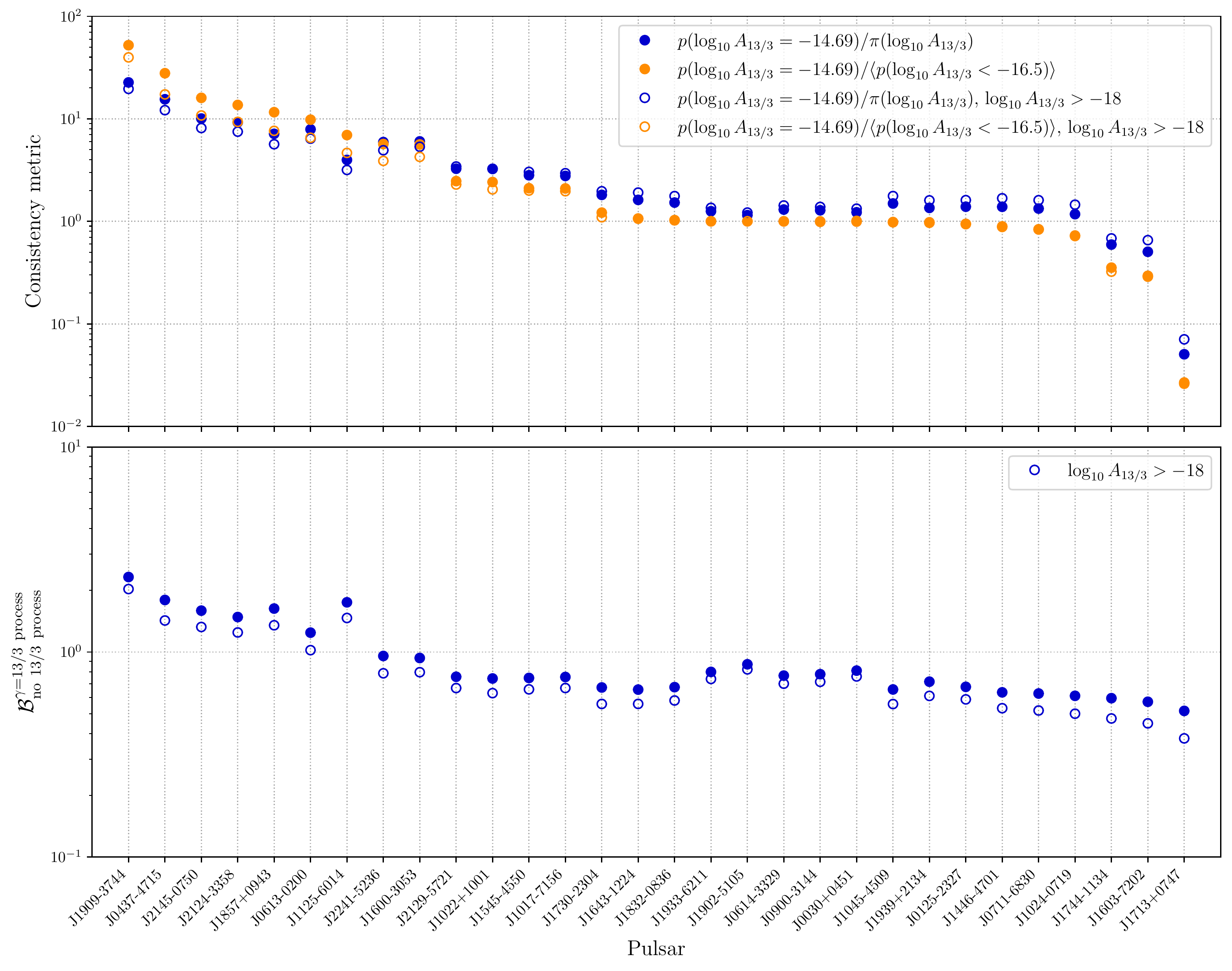}}
\caption{Statistics assessing the support for a noise process with $\gamma=13/3$ for each pulsar in PPTA-DR3. Top: posterior probability density ratio for a $\gamma=13/3$ process, at $\log_{10} A_{13/3}  = -14.69$, relative to $\log_{10} A_{13/3}  < -16.5$ (where the data are insensitive to a $\gamma=13/3$ process) in blue, and the prior density in orange. Bottom: The Savage--Dickey Bayes factor $\mathcal{B}$ for a $\gamma=13/3$ process at any amplitude, which is also the ratio of the orange to blue points in the top panel. Filled and hollow circles correspond to prior range lower bounds on $\log_{10} A_{13/3}$ of $-20$ and $-18$, respectively.}
\label{fig:support}
\end{figure*}

Next, we assessed the consistency of each of the pulsars with the common red process. Using the posterior distributions in Figure \ref{fig:factorised_likelihood}, we define metrics to assess the consistency of individual pulsars with the common process. We compute posterior probability density ratios, which compare the probability taken at the measured common amplitude $p(\log_{10}A_{13/3}  = -14.69)$, against both the prior probability density $\pi(\log_{10}A^{\mathrm{CRN}}_{13/3} )$ and the mean probability at low (effectively zero) amplitude $p(\log_{10}A_{13/3}  < -16.5)$. The probability ratios, shown in the top panel of Figure \ref{fig:support}, are intended to portray similar information to the ``dropout factors" used in previous works \citep[e.g.][]{NG12.5yrSGWB, PPTA_dr2_crn}. 
We also compute the Savage--Dickey Bayes factor, $\mathcal{B} = \pi(\log_{10}A_{13/3} ) / p(\log_{10}A_{13/3}  < -16.5)$, which provides the support from each pulsar for a $-13/3$ process at any amplitude (bottom panel in Figure \ref{fig:support}). The three pulsars with the highest $\mathcal{B}$ and the three with the lowest $\mathcal{B}$ are highlighted with unique colors in Figure \ref{fig:factorised_likelihood}.  The three pulsars that show the highest support show evidence for red noise in single-pulsar noise analyses and also are likely to dominate the factorized likelihood. The three pulsars with the lowest $\mathcal{B}$ show no evidence for red noise in single-pulsar analyses. The posterior distributions for these pulsars show negative support for the CRN at the inferred amplitude. This is discussed further below. 

\subsection{Common uncorrelated process}

\begin{figure*}
\centerline{\includegraphics[width=0.7\textwidth]{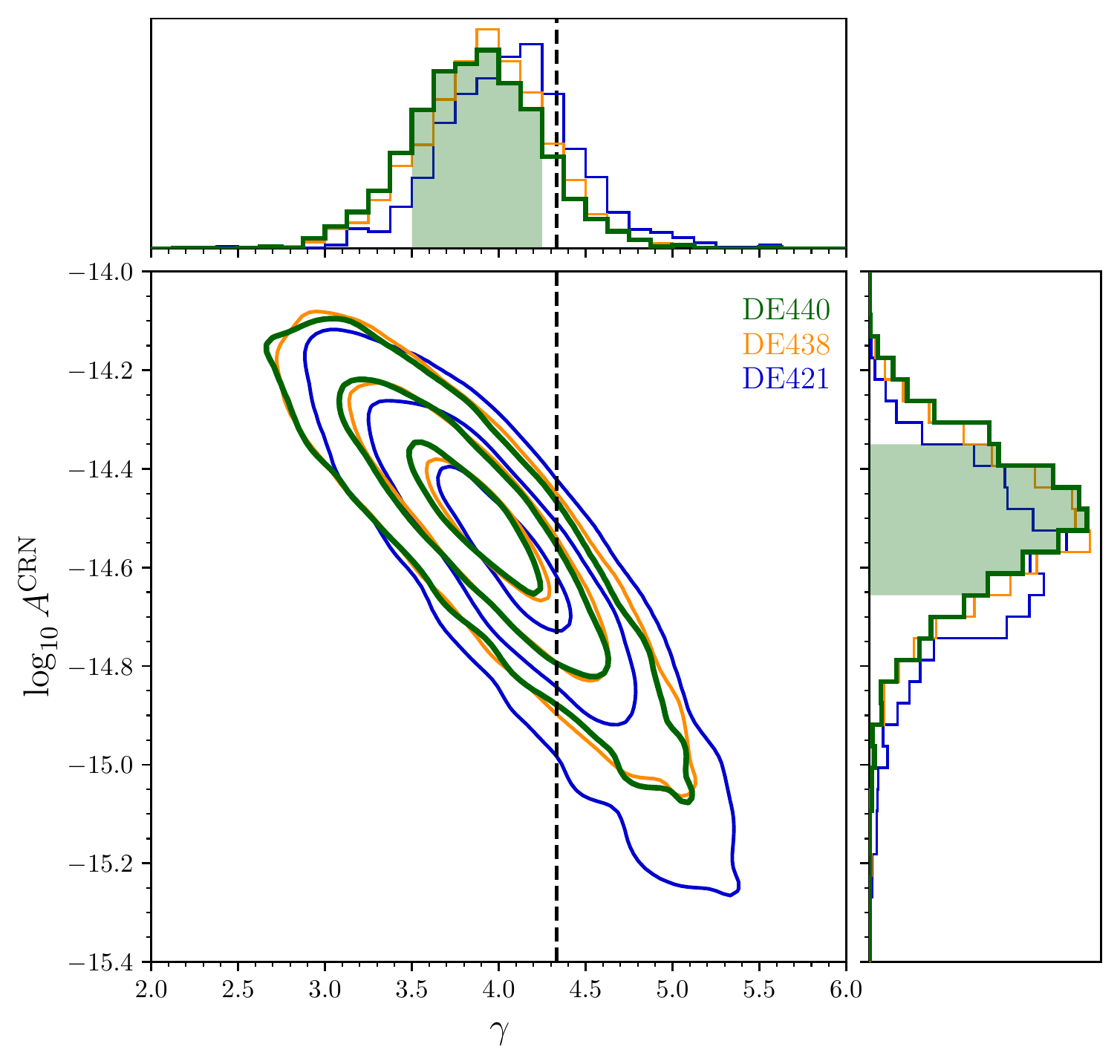}} 
\caption{Marginal posterior probability distributions for the measured logarithmic amplitude  and spectral index ($\gamma$) of a common uncorrelated process assuming different solar system ephemerides (SSE). Using DE440 (green) we measure $\gamma = 3.87 \pm 0.36$ and $\log_{10} A^{\mathrm{CRN}} = -14.50^{+0.14}_{-0.16}$ (median and 68\% credible interval; shaded regions in one-dimensional histograms). The contours on the two-dimensional marginal distribution show the 1$\sigma$, 2$\sigma$, and 3$\sigma$ credible intervals for each of the SSEs.}
\label{fig:curn_ephems}
\end{figure*}

We next search for a single uncorrelated common process with a variable spectral index. In contrast to the factorized-likelihood analysis discussed previously, this process is included for the PTA as a whole and sampled simultaneously with all red components of the single-pulsar noise models. The characteristics of this common process are considered for multiple SSEs from the Jet Propulsion Laboratory (JPL; DE421, DE438, and DE440), with our fiducial results generated with the most recent DE440 ephemeris \citep{DE440}. The one and two-dimensional marginal posterior probability distributions for the (log) amplitude and spectral index for the recovered process are shown in Figure \ref{fig:curn_ephems}. All ephemerides return consistent constraints for the process, with DE440 providing $\gamma = 3.9 \pm 0.4$ and $\log_{10} A^{\mathrm{CRN}} = -14.50^{+0.14}_{-0.16}$. While DE421 is an obsolete SSE that is inaccurate for the Jovian system, the recovered common noise is consistent, suggesting that the spectral characteristics are dominated by another source. Under the assumption of a fixed spectral index of $\gamma = 13/3$ we recover the same amplitude constraint as the factorized-likelihood analysis, as expected.

\begin{figure}
    \centering
    \includegraphics[width=\linewidth]{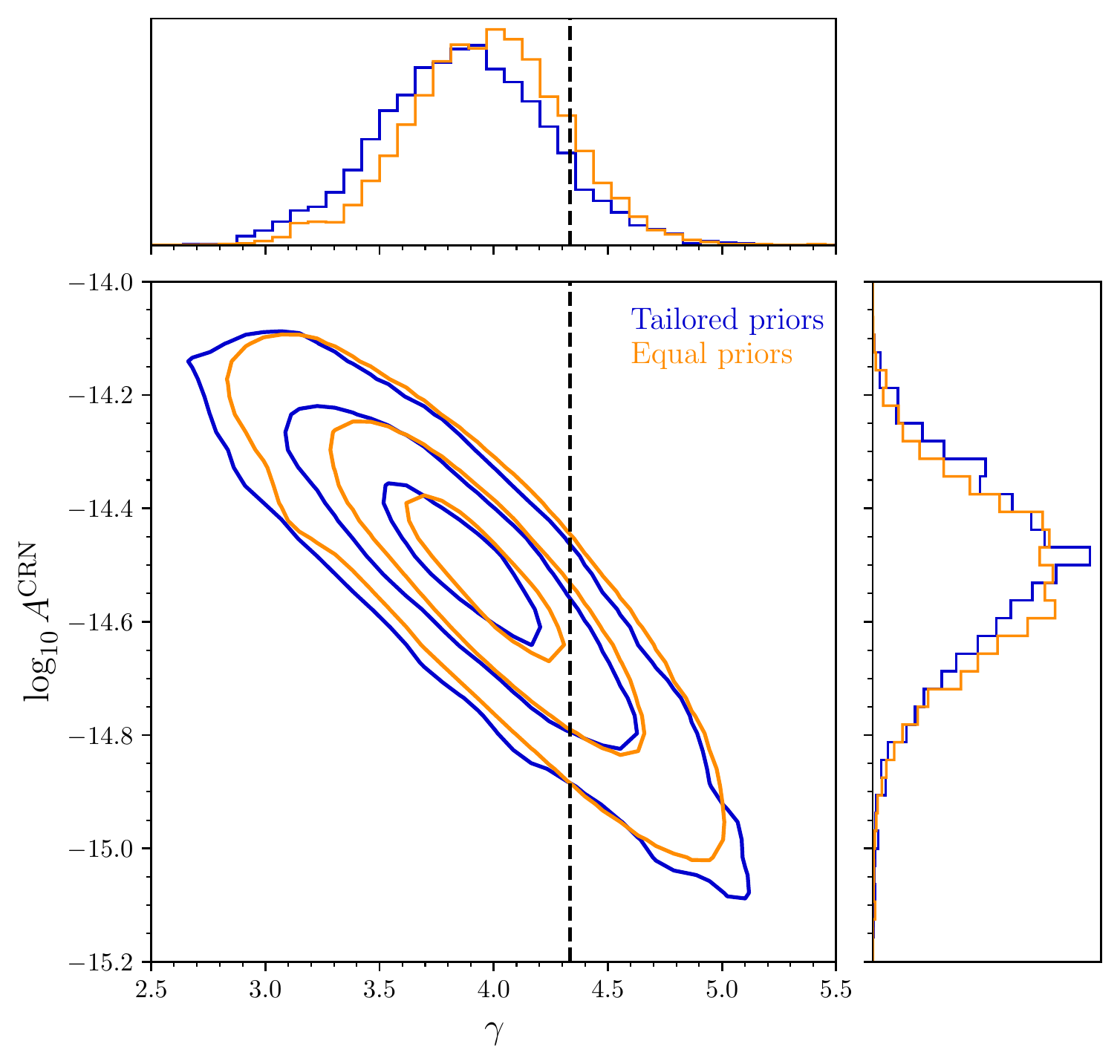}
    \caption{Marginal posterior probability distributions for the logarithmic amplitude and spectral index ($\gamma$) of a common uncorrelated process, under our tailored single-pulsar priors (blue) and broader equal priors (orange). }
    \label{fig:crn_prior}
\end{figure}

We also searched for a common uncorrelated process using commonly used broad priors on single-pulsar noise terms ($-20 \leq \log_{10} A \leq -11$,  $0 \leq \gamma \leq 7$) under the DE440 ephemeris. We compare this with our distribution recovered under tailored priors for single-pulsar noise terms in Figure \ref{fig:crn_prior}. The distributions are consistent, and for the broader priors we recover $\gamma = 4.0^{+0.3}_{-0.3}$ and $\log_{10} A^{\mathrm{CRN}} = -14.52^{+0.14}_{-0.15}$. The differences are comparable to those induced by SSE choice and much smaller than those induced by weakness in single-pulsar noise models \citep{PPTA-DR3_noise}. 

\subsection{Time-dependence of the common process amplitude}

\begin{figure}
\centerline{\includegraphics[width=0.5\textwidth, trim= 0 0 0 0 , clip]{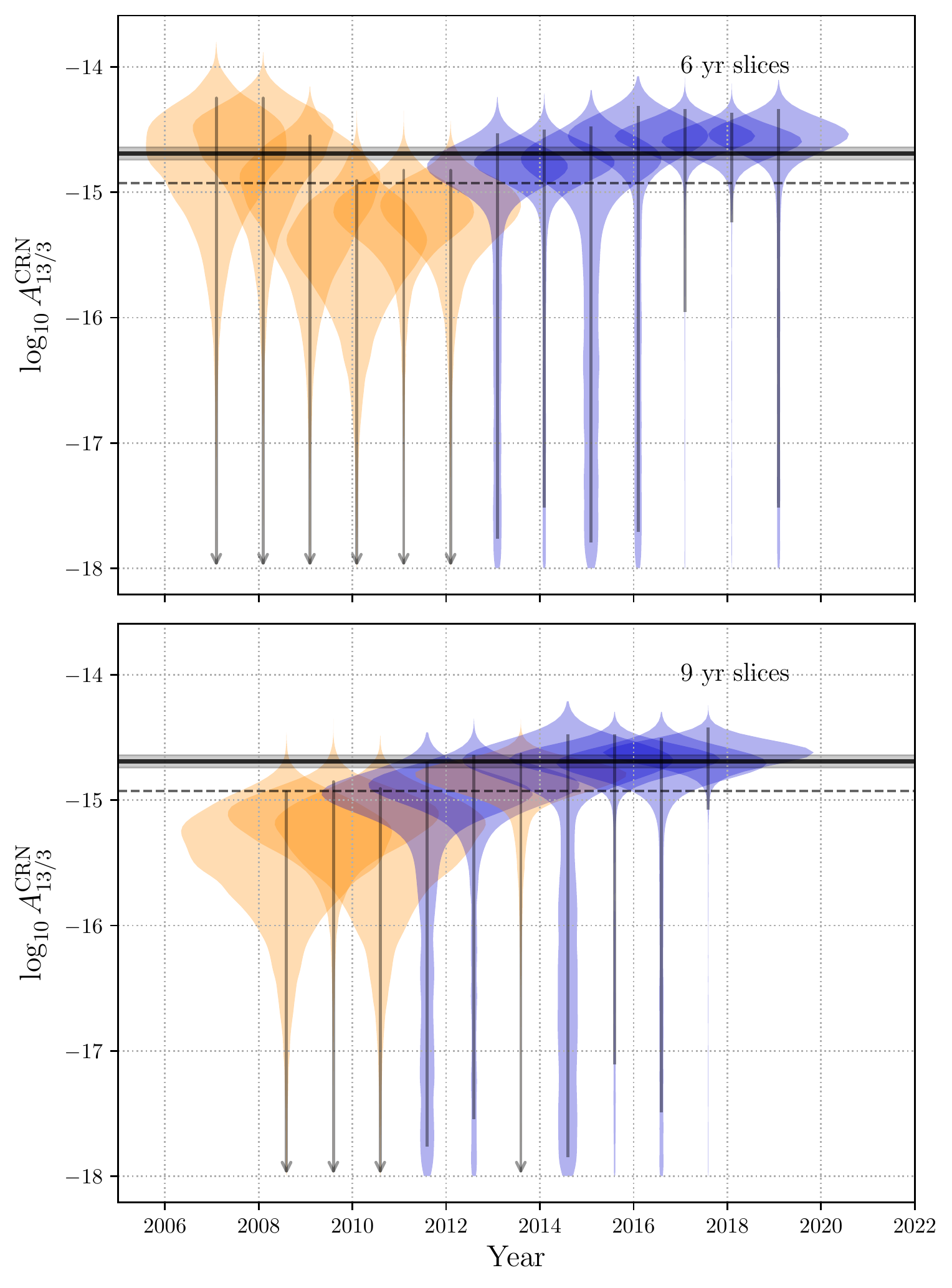}}
\caption{Posterior probability density \textit{violins} for the uncorrelated common-spectrum noise amplitude as a function of time, using a 6 yr (top) and 9 yr (bottom) sliding window (\textit{slices}) over the data set. The spectral index is fixed at $-13/3$. Slices with unconstrained (log) amplitude posteriors have been reweighted to have linear priors in $A^{\mathrm{CRN}}_{13/3}$ (orange), and the 95\% confidence upper limits are denoted with vertical arrows. Slices with constrained measurements of $\log_{10}A^{\mathrm{CRN}}_{13/3}$ are colored in blue. For reference, the dashed horizontal line indicates the $1.2\times10^{-15}$ upper limit set by our first 9 yr slice. The solid horizontal line and gray band indicate the measured $\log_{10 }A^{\mathrm{CRN}}_{13/3}$ of $-14.69$ and its 68\% credible interval from our full PPTA-DR3 analysis. Here and elsewhere, the violins represent the probability density of a parameter, with broader segments of a violin corresponding to higher probability density.}
\label{fig:time_slices}
\end{figure}

Next we search for any time evolution of the CRN process. 
With the longest multiwavelength timing baseline and high timing precision on many pulsars, the PPTA data provides the best opportunity to assess the stationarity of any purported common red signal. Our motivation is to investigate the reasons why the previous upper limits on the GWB from all PTAs were lower than the amplitude of the purported common noise being detected in recent and current data sets. This includes a limit of $A^{\mathrm{CRN}}_{\rm 13/3} < 1.2 \times 10^{-15}$ (95\% confidence), which we place using the first half of our data set (described below). This limit can be considered a replacement of the limit placed using a subarray of four PPTA pulsars with data extending to the beginning of 2015 \cite[][]{Shannon+15}, which used the now-obsolete DE421 SSE \cite[][]{DE421}.  

We estimate the amplitude of the common noise in 6 and 9 yr sliding windows (\textit{slices}) assuming $\gamma = 13/3$ and present the results in Figure \ref{fig:time_slices}. In the early time windows, marked in orange, a common-noise process with the $\gamma=13/3$ spectrum was not detected, and we have shown the 95\%-confidence upper limit for the amplitude. In the later windows, marked in blue, a detection was made, and the full 95\% credible interval was given. It is clear that these bounds on the common-noise increase with time, and in the earlier data, they are broadly consistent with earlier upper limits \citep[e.g.][]{Shannon+15, NG11yrSGWB}. The observational systems improved significantly with time, but it is hard to avoid the conclusion that, if the common noise represents a single distinct physical process such as a GWB, then the process is not time stationary. However, we cannot at this point rule out a nonastrophysical origin for this effect.

The earlier parts of the data used less powerful signal processing equipment, resulting in reduced sensitivity (and higher upper limits) due to lower received observing bandwidth and greater levels of digital artifacts in the data \cite[][]{Manchester+13, ppta_dr2}. In intermediate windows (dates centering between 2010 and 2012), which included data with the more sensitive last-generation narrowband (pre-UWL) digital receiving systems, we also find no evidence of a common red process. The 95\% upper limits were computed by reweighting the posterior samples to a prior that is uniform in linear amplitude. We infer upper limits from these intermediate windows ranging from $A^{\mathrm{CRN}}_{13/3} < 1.2\times 10^{-15}$ to $A^{\mathrm{CRN}}_{13/3} < 1.5\times 10^{-15}$, which are below the measured $A^{\mathrm{CRN}}_{13/3} = 2.0 \pm 0.2 \times 10^{-15}$ from the full data set. The upper limit from the first 9 yr ($A^{\mathrm{CRN}}_{13/3} < 1.2\times 10^{-15}$) slice corresponds with approximately the same data span and time range considered in \citet{Shannon+15}. The similarly tight bound that we derive consolidates this result and other similar upper limits \citep[e.g.][]{NG11yrSGWB}. The amplitude inferred from the most recent 6 yr slice is $A^{\mathrm{CRN}}_{13/3} = 2.7^{+0.8}_{-0.7} \times 10^{-15}$. This is higher than, but still consistent with, the $A^{\mathrm{CRN}}_{13/3}$ measurement from the full data span. \citet{Johnson+22} found that the use of a small number of pulsars \citep[as in][]{Shannon+15} can increase the chance of underestimating upper limits on the amplitude of a common process. However, this is not the origin for the apparent time variability that we observe, as we use all available pulsars in each slice (i.e. 23 pulsars for the first slice, and 30 pulsars in the last) and the same priors and noise models across the slices.

Using the first three 9 yr windows, the upper limits on the amplitude of the signal are inconsistent with the amplitude measured in the entire data set, and in later subsets. The measured uncorrelated common-spectrum amplitude $\log_{10} A^{\rm CRN}_{13/3} = -14.69$ lies at the 99.8 percentile of the reweighted samples from the first 9 yr window. While some variability would be expected because of the stochasticity of the background \cite[][]{Hazboun+20a}, this level of variation is extreme, and the implications are discussed below. 

\subsection{Common free spectrum}

\begin{figure*}
\centerline{\includegraphics[width=0.8\textwidth, trim= 0 0 0 0 , clip]{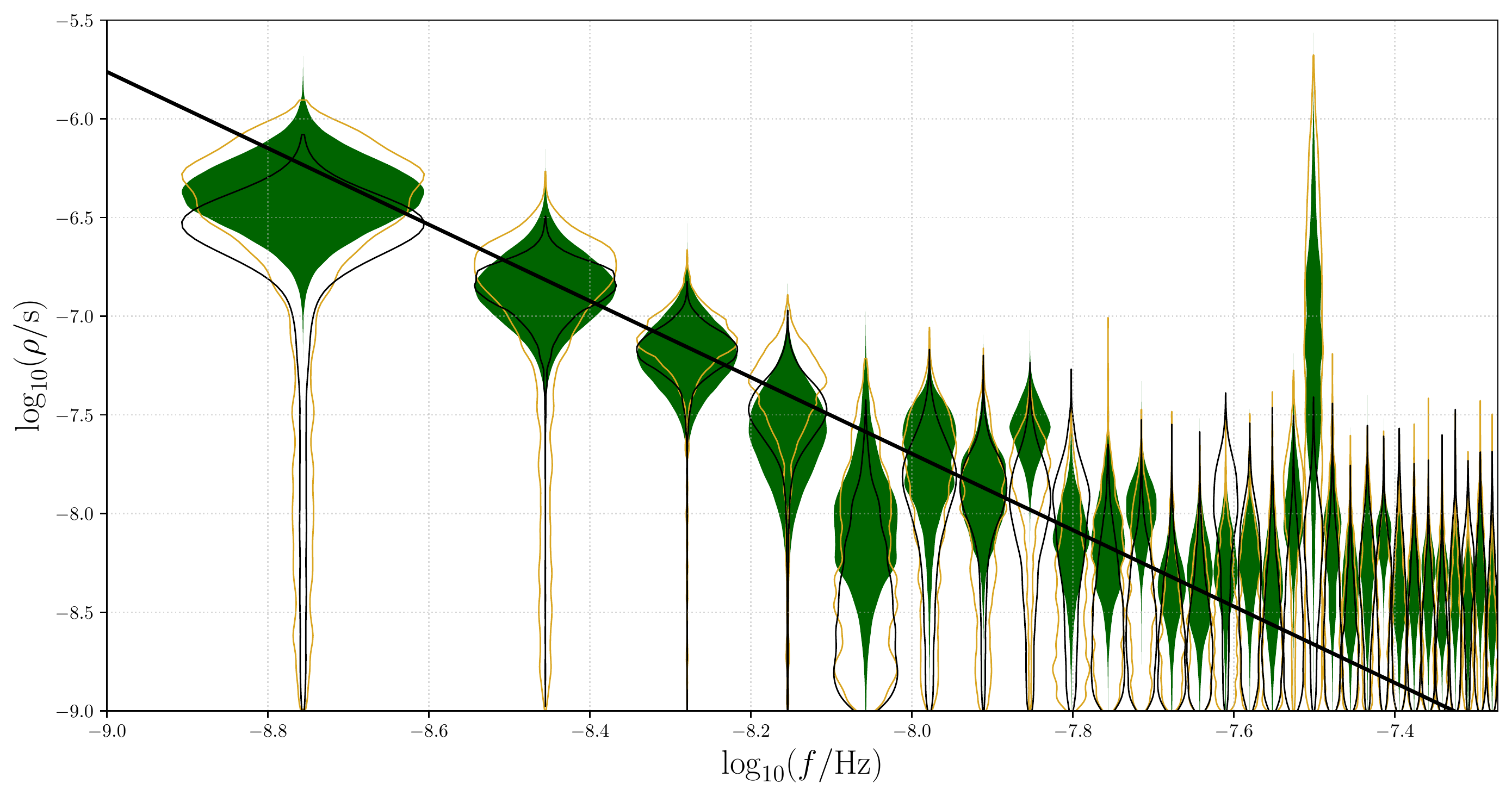}}
\caption{Free-spectrum common-noise inference for the PPTA-DR3. The textit{violins} show the probability density of the power spectral density for the free spectrum of a common-noise process (the width of the violin represents the probability density, with a linear scale). Solid green violins are from a factorized-likelihood analysis using the achromatic noise in single pulsars. Black violins are from a common free-spectrum inference assuming no cross correlations. Gold violins show the free spectrum of a common-noise process with Hellings--Downs cross correlations.}
\label{fig:freespec}
\end{figure*}

To assess whether or not the apparent common red signal is consistent with a power-law process, we formed the free spectrum. As with power-law processes, the free spectrum is modeled as a Fourier series but the variances of each Fourier coefficient are independent (i.e., not constrained to follow a power-law process). 

We formed the free spectrum using multiple methods. First we modeled it as a common process both with zero correlations and with Hellings--Downs correlations. In addition, we formed the spectrum using a factorized-likelihood approach. The probability density for the PSD of individual Fourier components can be factorized in the same way as the power-law amplitude. We determine, for each pulsar, the probability density of the (achromatic) noise PSD for each Fourier component. The frequencies for the components are the same for each pulsar, and are the frequencies that the total dataset is sensitive to, from $1/T_{\rm span}$ to $60/T_{\rm span}$, where $T_{\rm span}$ is the total length of the PPTA-DR3 dataset.

The different estimates of the free spectrum can be seen in Figure \ref{fig:freespec}. The results are consistent, showing that the common noise is well described as the probability-weighted mean of all achromatic noise processes in individual pulsars. Additionally, it is consistent with a power law, particularly at the lowest frequencies, which suggests that a single process may be dominating. 
The factorized-likelihood spectrum shows excess power at a frequency of 14.0\,nHz, which corresponds to a period of 2.26\,yr. The common free-spectra show reduced sensitivity for periods near 3 yr, which corresponds to the approximate time span of the new pulsars added to the array (with only UWL data). Including Hellings--Downs correlations does not significantly change the nature of the spectrum. We interpret this as a result of the spectral characteristics of the common noise being dominated by the autocorrelations \citep{Spiewak+22}.

\subsection{Isotropic Stochastic Gravitational-wave Background}

We searched for evidence of the correlations of an isotropic GWB by considering models that include the Hellings--Downs spatial correlations in the PTA likelihood. We first searched for a power-law process with an unknown spectral index and the DE440 SSE, recovering $\gamma = 3.87 \pm 0.47$ and $\log_{10} A^{\rm HD} = -14.51^{+0.18}_{-0.20}$, which is consistent with, but less precise than the measurement using the uncorrelated component alone. The agreement is unsurprising as the sensitivity of the PPTA data set is dominated by the autocorrelations. At a fixed spectral index for the assumed GWB from SMBHBs, we measure $\log_{10} A^{\rm GWB} = -14.68 \pm 0.06$, which is also consistent with the uncorrelated noise. We have also measured the common-noise free spectrum assuming Hellings--Downs correlations and find that it is consistent, at all Fourier frequencies considered, with the uncorrelated common noise (Figure \ref{fig:freespec}, gold). Again this indicates that the cross correlations are dominated by the autocorrelated component.

To quantify whether the data supports the inclusion of these correlations, we estimate the Bayesian odds ratio for using product-space sampling \citep{NG11yrSGWB, NG12.5yrSGWB}. Relative to a common uncorrelated process, we find no additional Bayesian support, for or against, the Hellings--Downs using this approach, with $\mathcal{B}^{\rm HD}_{\rm CRN} \sim 1.5$ for a varied spectral index, and $\mathcal{B}^{\rm HD}_{\rm CRN} \sim 2$ for $\gamma = 13/3$. These values are not significant.

\subsection{Search for the spatial correlations}

We directly search for the spatial correlations using our technique described in Section \ref{sec:corr_method}, instead of assuming an ORF and testing for model support through a common-noise model. The measurement of interest here is the correlation coefficient $\Gamma$ of a $\gamma = 13/3$ process, and we obtain one measurement (the posterior probability density) for each of the 435 unique pulsar pairs in our array. These measurements are informed by the cross correlations only and assume a common distribution for $\log_{10} A_{13/3}$. The parameters in the single-pulsar noise models have also been accounted for in each pair. The numerically marginalized single-pulsar noise properties are self-consistent, meaning that the observed noise characteristics of a pulsar are independent of which pulsar it is paired with.

\begin{figure*}
\centerline{\includegraphics[width=0.8\linewidth]{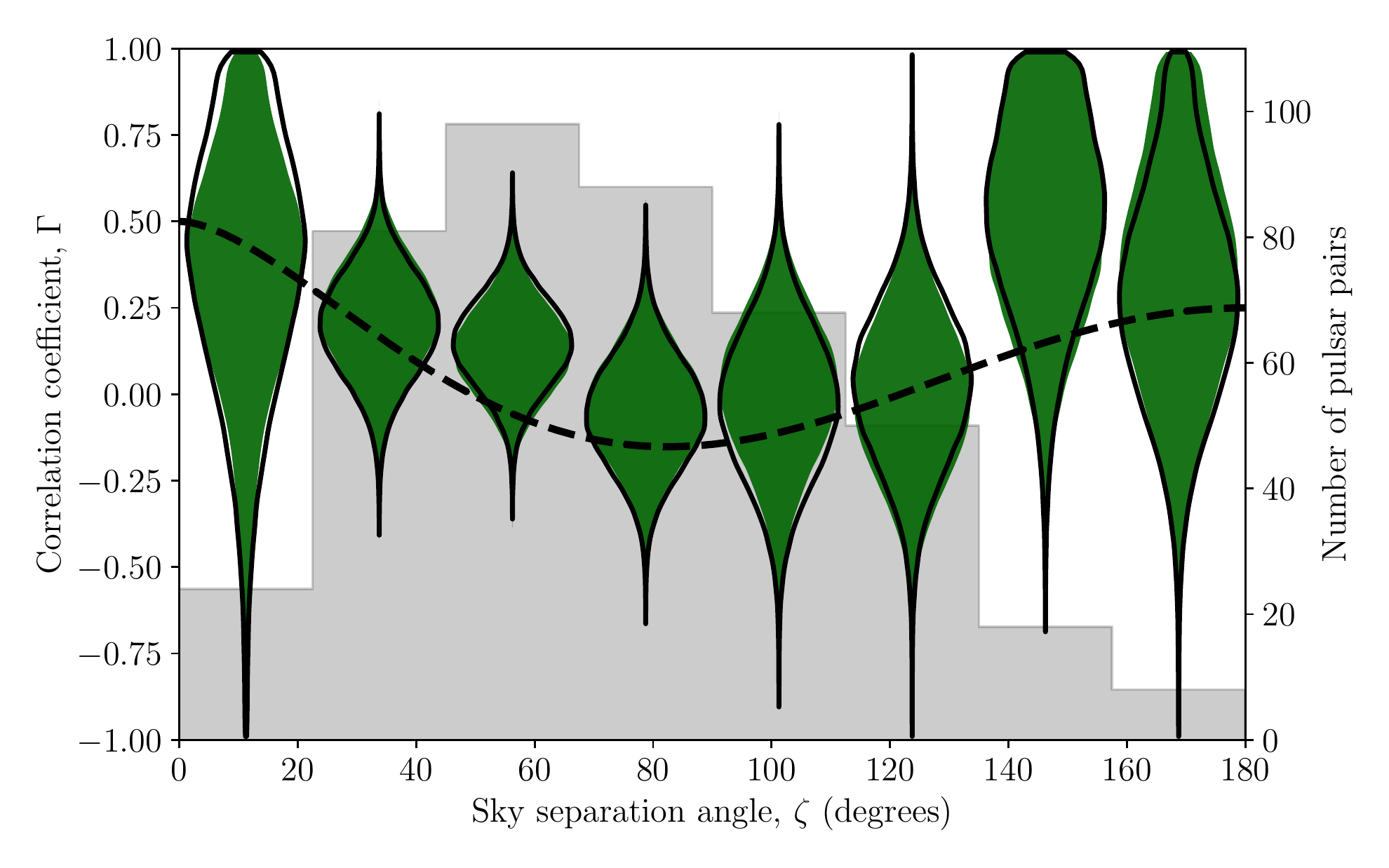}}
\caption{Measured spatial correlations as a function of the angular separation angle, $\zeta$. The width of each \textit{violin} as a function of $\Gamma$ is proportional to the inferred probability density, $p(\Gamma)$. Binned correlations are shown assuming a common-noise distribution of $\log_{10} A^{\mathrm{CRN}}_{13/3} = -14.69 \pm 0.05$ (green, filled) and with amplitude fixed at the median (black, hollow). The dashed black line is the Hellings--Downs ORF. The gray histogram shows the number of pulsar pairs in each angular bin.}
\label{fig:correlations}
\end{figure*}

To visualize the correlations, we place our 435 pulsar-pair PDFs, $p(\Gamma)$, into eight independent and equally spaced bins in sky separation angle $\zeta$. The PDFs in each bin are multiplied to give the total probability density for that bin. The resulting PDFs are shown in Figure \ref{fig:correlations}, with a gray histogram representing the number of pulsar pairs in each angular bin. It should be noted that because the cross-correlation measurements are not completely independent, the apparent uncertainties will be underestimated (emphasizing the need for statistical validation using bootstrap noise realizations).

We show both the measurements assuming a common distribution of $\log_{10} A_{13/3}  = -14.69 \pm 0.05$, and assuming a fixed amplitude of $\log_{10} A_{13/3}  = -14.69$. The measurements were produced with independent posterior samples and the resulting correlations are nearly identical, which indicates that the effect of KDE uncertainty due to a finite number of samples is negligible. It can be clearly seen that the Hellings--Downs curve (solid black line, Equation \ref{eqn:hd}) is a better description of the resulting PDFs than zero correlations (which we quantify below). It can also be seen that a monopole (with correlation coefficient, $\Gamma = 1$) and dipolar correlations (with $\Gamma = \cos(\theta)$, i.e.,  $\Gamma=1$, at $\zeta = 0^\circ$, and $\Gamma = -1$, at $\zeta = 180^\circ$) are not the dominate source of the common noise at this assumed amplitude and spectral index. However, these alternative correlations must be present at some level in all PTA datasets.

\begin{figure}
\centerline{\includegraphics[width=0.5\textwidth, trim= 0 0 0 0 , clip]{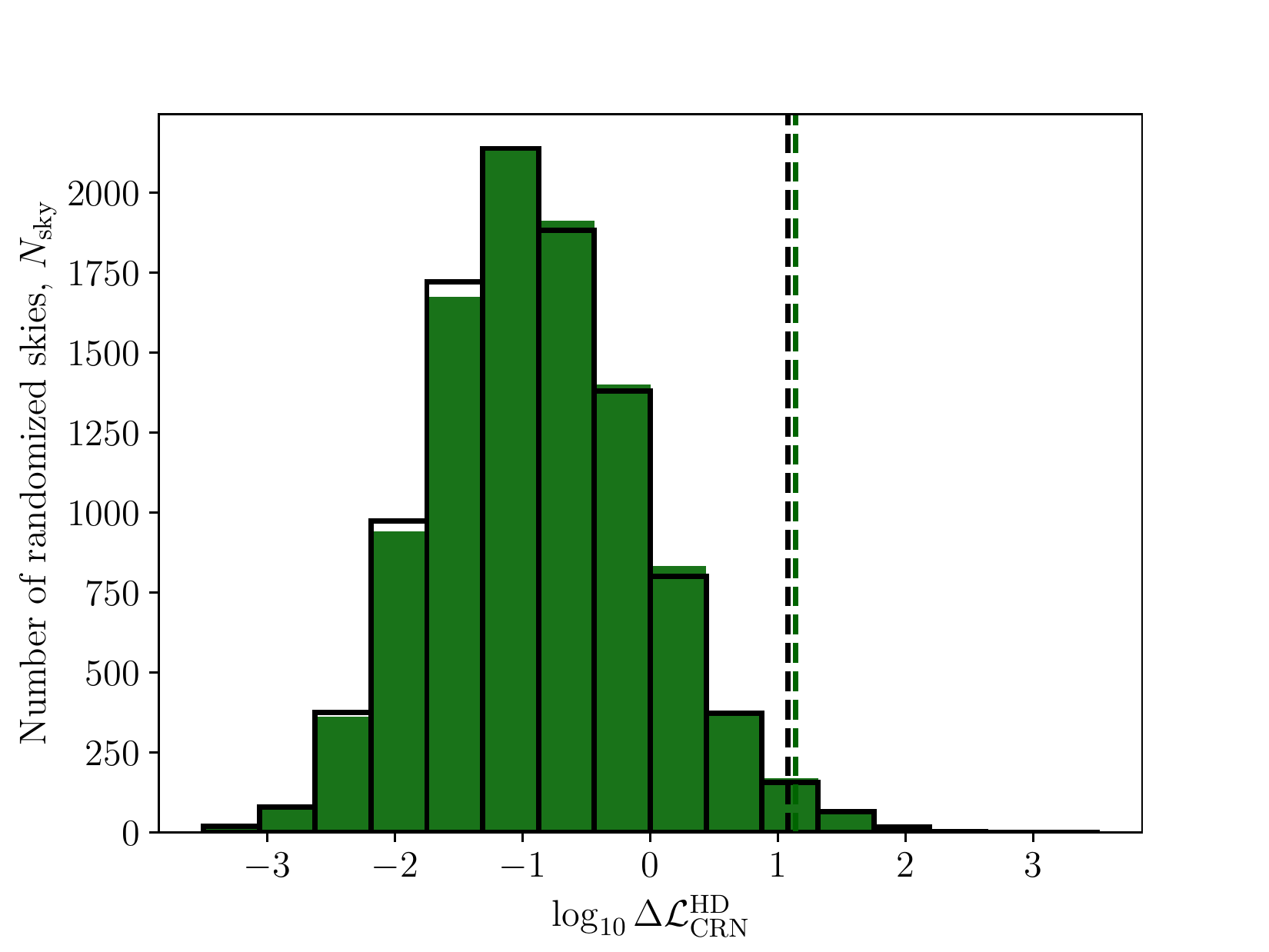}}
\caption{False-alarm probability calculations from $10^4$ quasi-independent randomized pulsar sky distributions (sky scrambles). As with Figure \ref{fig:correlations}, the green (filled) histogram corresponds to the measurements assuming $\log_{10} A^{\mathrm{CRN}}_{13/3} = -14.69 \pm 0.05$ and the black (hollow) histogram assumes the median amplitude. The measured likelihood ratios $\Delta\mathcal{L}^{\rm HD}_{\rm CRN}$ from the data under the two assumptions about the amplitude are marked by dashed lines and correspond to a one-sided false-alarm probability of $p \lesssim 0.014$.}
\label{fig:scambles}
\end{figure}

We compute the model likelihoods for the correlations associated with the common noise and find a likelihood ratio $\log_{10}\Delta\mathcal{L}^{\rm HD}_{\rm CRN} = 1.1$ in support of Hellings--Downs over zero correlations. To interpret the significance of this result, we calibrate the false-alarm probability by generating noise from the $p(\Gamma)$ distributions themselves via sky scrambling. We generate $10^4$ random pulsar sky distributions with the criteria for near-independence described in Section \ref{sec:corr_method}. As the sky distributions are not truly independent, the false-alarm probability is only an estimate \citep[for discussion see ][]{dimarco+23}. 
A histogram of the $\log_{10}\Delta\mathcal{L}^{\rm HD}_{\rm CRN}$ for Hellings--Downs over uncorrelated noise from these randomized skies is shown in Figure \ref{fig:scambles}, from which we derive one-sided $p$-values of $p \lesssim 0.014$ and $p \lesssim 0.015$ for our observations from the varied and fixed-amplitude correlations, respectively. Using Silverman's rule \citep{Silverman86} to set the kernel bandwidth for each pair, we observe consistent values of $p \lesssim 0.018$ and $p \lesssim 0.012$, respectively. We note that, while sky scrambling with this method is simple and efficient, other false-alarm probability calculations will be required for validating any future GWB detection claims \citep[e.g. phase shifting and accounting for the different sensitivities of the pulsars; ][]{Taylor+17}.

\begin{figure*}
\centerline{\includegraphics[width=0.95\textwidth, trim= 0 0 0 0 , clip]{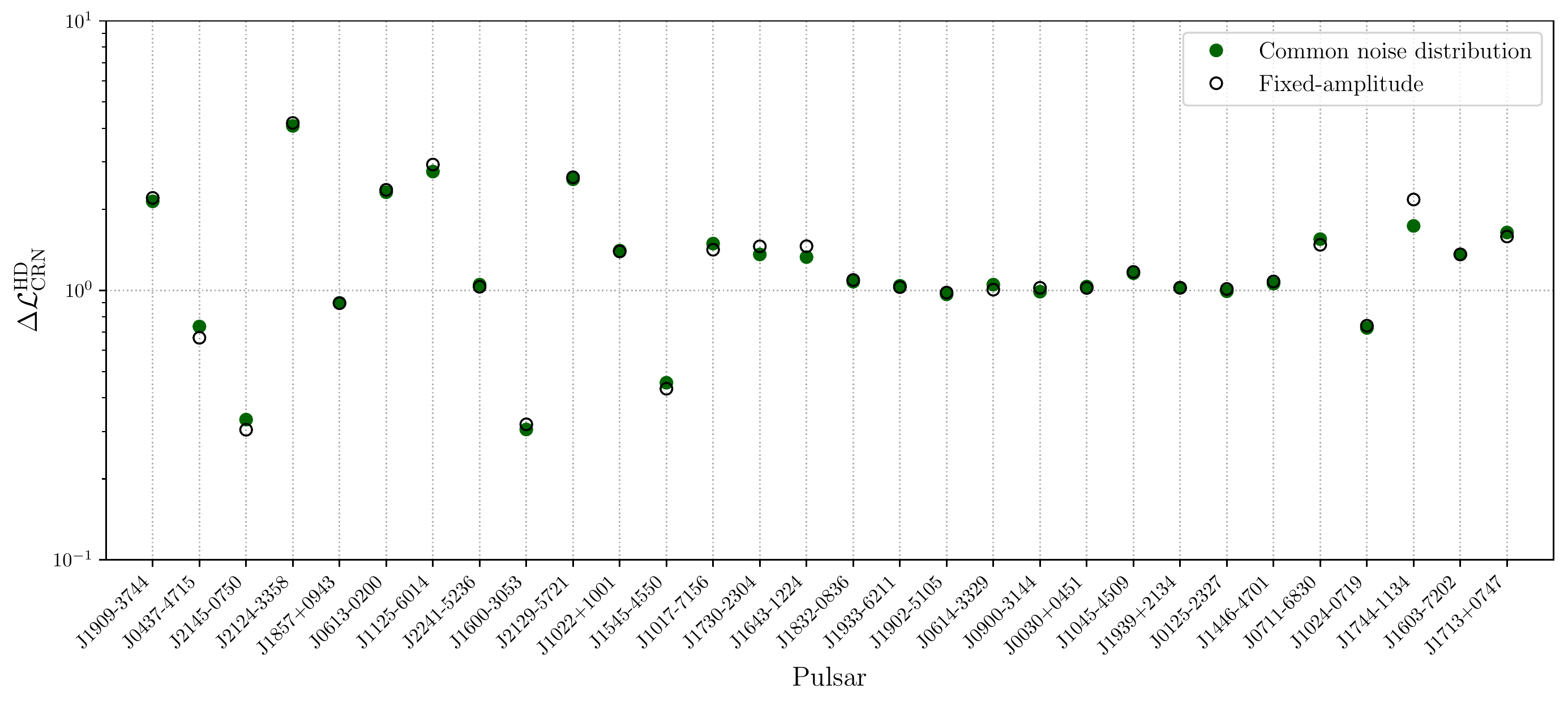}}
\caption{Likelihood ratio for Hellings--Downs (HD) spatial correlations over zero correlations for a $\gamma=13/3$ process, at an amplitude of $\log_{10} A_{13/3}  = -14.69$. Each point is calculated using the 29 pulsar pairs involving the labeled pulsar, and therefore, the points are not independent.}
\label{fig:hdsupport}
\end{figure*}

The support for Hellings--Downs correlations can be efficiently quantified on a per-pulsar basis with our scheme. Figure \ref{fig:hdsupport}  shows, for each pulsar, the likelihood ratio (Bayes factor estimator) $\Delta\mathcal{L}^{\rm HD}_{\rm CRN}$ computed using only pairs involving the given pulsar. The pulsars are ordered by their support for the uncorrelated common noise in Figure \ref{fig:support}. The pairs involving PSR~J1744$-$1134 give the most discrepant $\Delta\mathcal{L}^{\rm HD}_{\rm CRN}$ between the two amplitude assumptions we consider. This pulsar has fewer independent posterior samples for the varied amplitude analysis than most others, because of its low likelihood support in this region of the parameter space (PSR~J1713$+$0747, with lower likelihood support, was analyzed including samples from additional MCMC chains to compensate). Therefore the difference likely reflects some small KDE uncertainty due to a finite number of samples. 

We compute the number of effective pulsar pairs using $\sigma_k$ from each pulsar pair $k$ using Equation \ref{eqn:neff}, and find $n_{\rm eff} = 355$.

\subsection{Marginalizing over solar system ephemeris errors}

An accurate model for the position of the center of the solar system is paramount for the timing precision required by PTA experiments. Systematic errors in the positions or masses of the planets can add additional signals to pulsar timing data sets that could manifest as both spatial correlations, but also as a common-noise process. The most recent SSE from JPL, DE440, used data from the Juno mission and updated very-long-baseline interferometry measurements of the outer planets, and is described as having reduced systematic uncertainties \cite[][]{DE440}. 

We investigated if the amplitude of the apparent CRN could be affected by systematic uncertainties by perturbing the masses of outer planets and the orbital elements of the Jovian system with \textsc{Bayesepehem} models \cite[][]{Bayesephem}, with respect to the ephemerides DE421, DE438, and DE440. 
We find that in all ephemerides considered, our data recovers a nonzero (95\% confidence) perturbation to at least one orbital element of the Jovian system. We note that there was no Bayesian evidence for perturbations in the DE438 ephemeris using PPTA-DR2 \citep{PPTA_dr2_crn}. 

\begin{figure*}
\centerline{\includegraphics[width=0.7\textwidth, trim= 0 0 0 0 , clip]{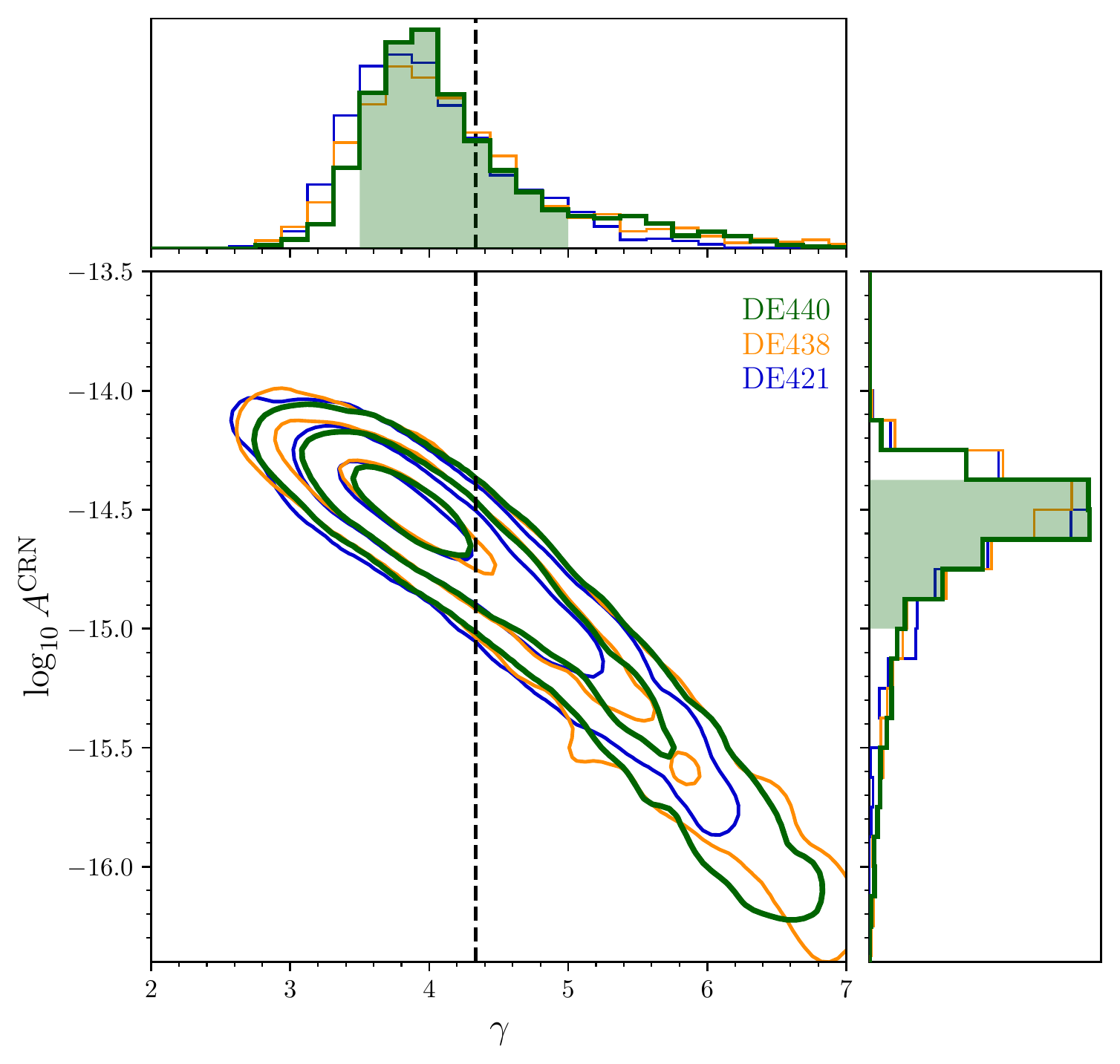}}
\caption{As with Figure \ref{fig:curn_ephems}, but with \textsc{Bayesephem} used to marginalize over potential solar system ephemeris errors. We measure $\gamma = 4.02^{+ 0.81}_{- 0.41}$ and $\log_{10} A^{\mathrm{CRN}} = -14.56^{+ 0.16}_{- 0.39}$ for DE440 (green).}
\label{fig:bayesephem}
\end{figure*}

If we marginalize over these potential perturbations, we recover a steeper spectral index for the common-noise process, $\gamma = 4.02^{+ 0.81}_{- 0.41}$, with a log amplitude $\log_{10} A^{\mathrm{CRN}} = -14.56^{+ 0.16}_{- 0.39}$ with DE440. All ephemerides considered give consistent constraints for this process, as shown by the posterior densities in Figure \ref{fig:bayesephem}. While we observe weak evidence for SSE errors in the form of nonzero perturbations, it is assumed that if such errors were a significant source of common noise, they would manifest as a dipole-like feature in the inter-pulsar correlations \citep{Tiburzi+16}. Our data shows that the measured common-noise process is not dominated by dipolar correlations, and we have therefore assumed that the DE440 SSE is sufficiently accurate for the purpose of our GWB search.

\section{Discussion}
\label{sec:discussion}
 
The properties of the common noise are consistent, within reported uncertainties, with measurements from recent works by the PPTA and the constituent members of the IPTA. A previous analysis by the PPTA \citep{PPTA_dr2_crn} found a common uncorrelated process with $\gamma = 4.11^{+ 0.52}_{- 0.41}$ and $\log_{10} A^{\mathrm{CRN}} = -14.55^{+ 0.10}_{- 0.23}$. The EPTA \citep{EPTA_dr2_crn} have reported a similarly consistent result of $\gamma = 3.78^{+ 0.69}_{- 0.59}$ and $\log_{10} A^{\mathrm{CRN}} = -14.29^{+ 0.25}_{- 0.33}$. NANOGrav \citep{NG12.5yrSGWB} identified a common-spectrum process with a steeper median spectral index, remaining consistent through relatively larger uncertainties reported with the value $\gamma = 5.52^{+ 1.26}_{- 1.76}$ and $\log_{10} A^{\mathrm{CRN}} = -14.71^{+ 0.14}_{- 0.15}$. This work used only the five lowest-frequency components, and a shallower process was recovered when more components were considered. This may be due to high-fluctuation-frequency processes present in the data set, such as unmodeled pulse-shape variations \cite[][]{Shannon+16, NG12.5yrSGWB, PPTA_dr2_noise}.

The search for a common-spectrum process through the combination of a previous generation of these data sets by the IPTA \citep{IPTA_dr2_gwb} also finds a consistent signal most closely resembling that identified by the EPTA, $\gamma = 3.90^{+ 0.90}_{- 0.90}$ and $\log_{10} A^{\mathrm{CRN}} = -14.29^{+ 0.14}_{- 0.36}$. The PPTA used the DE436 SSE for the previous analysis, while other PTAs used DE438. The spectral index marginally steepens when we marginalize over the SSE errors modeled by Bayesephem, and becomes more consistent with the prediction for the background predicted by SMBHBs. 

This common spectral noise may be caused by the GWB, or it may represent one or more other signals masquerading as a background \citep{Tiburzi+16,Zic+22}. If the purported common-spectrum noise is a genuine signature of an isotropic, stochastic GWB, then we can expect the IPTA to make a detection using the current-generation data sets \citep{Siemens+13}.  In this scenario we need to convincingly explain (1) why three of our pulsars offer negative likelihood support for the existence of the GWB, (2) why the current detected amplitude is higher than previous upper bounds, and (3) why the apparent amplitude of the GWB in our data is measured to be increasing with time.

There are several reasons why individual pulsars may not exhibit the signature of the GWB at the same amplitude. This includes interaction between the GWB signal and intrinsic pulsar timing noise (although the probability of this decreases as data spans increase), misspecification of the intrinsic noise parameters in the modeling, errors in the pulsar timing model, or natural variance in the SMBHB source distribution and anisotropy. 

In the PPTA-DR3 data set, the pulsars most discrepant with the inferred common noise are PSRs~J1744$-$1134, J1603$-$7202, and J1713$+$0747. Two of these (PSRs~J1744$-$1134 and J1713$+$0747) are observed by all of the constituent member PTAs of the IPTA, and hence it is key to determine whether they show positive or negative support for the GWB in other data sets, and to explain any discrepancies. Previous analyses have revealed similar effects, including PSR~J1713$+$0747, which exhibited low dropout factors in the previous NANOGrav GWB search analysis \citep{NG12.5yrSGWB}. This may be attributed in part to imperfect modeling of the known timing events observed in this pulsar \citep{Lam+18}. PSR~J1603$-$7202 is observed as part of the MeerKAT PTA \citep{mpta}, which has shorter, but significantly more sensitive measurements of this pulsar at greater cadence. The MeerKAT PTA has fortnightly observations and should be sensitive to the noise processes at higher fluctuation frequencies.

The time-dependence of the signal amplitude is puzzling under an isotropic GWB model. Limits placed on the amplitude of the signal in earlier subsets of the data are inconsistent with high probability ($\log_{10} A_{13/3} < -14.69$ with $99.8\%$ confidence) with the amplitudes of the signal measured in later data, and the data set in the entirety. As shown in Figure \ref{fig:time_slices}, the amplitude obtained with the entire data set is consistent with that from the most recent 9\,yr data span. This result is heavily influenced by the most dominant pulsar, PSR~J1909$-$3744. The noise characteristics of this pulsar are unusual, and reflect the recovered properties of the common-spectrum noise. When considering the entire data set, there is only modest evidence for a deviation in the common-spectrum process from a pure power law.

If our results do not represent the signal from a GWB, then we must consider the implications.  The three  pulsars that offer negative likelihood support for the existence of a $\gamma = 13/3$ process at the measured common amplitude could therefore be representing an approximate upper bound on the amplitude of the background. The amplitude and spectral slope of the GWB is poorly constrained, and so lower amplitudes are still plausible. We reiterate the findings of the previous PPTA analysis \citep{PPTA_dr2_crn} that the characterization of the common noise can be affected by misspecification, and it is not necessarily a noise floor. That is, the sensitive pulsars with zero red noise can be outweighted in the likelihood by a population of pulsars with similar (but not necessarily identical) noise characteristics \citep{Zic+22}. As it is currently defined, the measured common noise is consistent with the probability-weighted mean of the achromatic noise in our sample of pulsars, across all frequency components. 

In the PPTA-DR3 data set, the number of pulsar pairs is not highly sensitive to the correlated component of the GWB at the amplitude of the purported CRN. Using autocorrelations and cross correlations, we find no Bayesian support, for or against, the presence of a GWB.  However, using our technique that hierarchically analyses the correlations of the common noise alone, we find a $\log_{10} \Delta\mathcal{L}^{\rm HD}_{\rm CRN} = 1.1$ in support of the Hellings--Downs over zero correlations. We computed the significance of this likelihood ratio via sky scrambling, which provides a reliable false-alarm probability through noise generated with the data itself.  It is unclear why the latter method provides greater support than the former. It could be that the former method requires the autocorrelated and cross-correlated noise to have the same amplitude. If the pulsar noise is misspecified, a mismatch in the amplitude of these two terms could degrade the sensitivity of the statistic. From this result, we conclude that future GWB searches should use (Bayesian) statistical methods that search for the GWB in the cross correlations alone \citep[e.g.][]{NG12.5yrSGWB}. Confirmation of consistency of the amplitude in the cross correlations and autocorrelations will be necessary to build confidence in using the entire PTA signal for astrophysical interpretation. 

\section{Conclusions}
\label{sec:conclusions}

We have presented a search for an isotropic stochastic gravitational-wave background using the third data release of the PPTA. We have measured the characteristics of a common-spectrum process in the array. Using the pulsar autocorrelations only, we measure $\log_{10}A^{\mathrm{CRN}} = -14.5$ and $\gamma = 3.87$, when using the DE440 SSE. By marginalizing over potential errors in the SSE, we find that the inferred process steepens with $\log_{10}A^{\mathrm{CRN}} = -14.56$ and $\gamma = 4.02$, although there is only weak evidence for these errors. At a fixed spectral index of $\gamma=13/3$ we find $\log_{10}A^{\mathrm{CRN}}_{13/3} = -14.69 \pm 0.05$, which translates to a constraint on the linear strain amplitude of $A = 2.04^{+0.25}_{-0.22}\times 10^{-15}$ for the model of an isotropic GWB. By analyzing only the spatial correlations using a new technique, we show that the process at this amplitude is consistent with an isotropic GWB at a level that may occur by chance with a probability of $p \lesssim 0.014$. This false-alarm probability corresponds to a one-tailed test significance of approximately $2\sigma$.

However, some of the apparent characteristics of this common-spectrum process are surprising under the model of an isotropic GWB, including the following: 
\begin{itemize}
    \item {\bf Three pulsars disfavor the presence of a $\mathbf{\bm{\gamma} \bm{=}13\bm{/}3}$ process at the measured \textit{common} amplitude.} The measured common-spectrum process, as it is defined, is not necessarily a \textit{noise floor} as would be expected of a GWB. We have demonstrated that the spectral characteristics of the common noise are nearly identical to the probability-weighted mean of all achromatic pulsar noise, which may have significant contributions from intrinsic spin noise. However, these same pulsars appear to contribute positively to the Hellings--Downs correlations when the measured common-noise process is introduced, which may suggest misspecification in the noise models of the pulsars.
    
    \item {\bf The apparent amplitude of the common noise changes as a function of time.} This observation cannot be explained simply by the growing sensitivity of the data. For approximately the first half of our dataset, we appear to be in a nondetection regime, where only upper limits can be placed on the amplitude of a common-spectrum process at $\gamma = 13/3$. The 95\% confidence upper limits derived from these early parts of our dataset are in tension with the current measurements. For example, in the first 9 yr slice, we find $\log_{10} A^{\mathrm{CRN}}_{13/3} < -14.92$ at $95\%$ confidence and $\log_{10} A^{\mathrm{CRN}}_{13/3} = -14.69$ is ruled out at $99.8\%$ confidence. 
    
    Similar results have been observed by NANOGrav and the PPTA during previous GWB searches. However, the significance of this tension has been debated, owing to differences in the data sets (e.g. differing numbers of pulsars) and analysis procedures (e.g. choice of SSEs, priors, and single-pulsar noise models). For this analysis, we used identical models, priors, and inference techniques on all subsets of the data. We cannot determine the origin of this apparent nonstationarity as the recovered common-noise amplitude (measured as a function of time) is not necessarily determined by a distinct physical process.
    
    \item {\bf There is no Bayesian support, for or against, the Hellings--Downs spatial correlations using both the autocorrelations and cross correlations.} Yet by analyzing only the cross correlations, using our computationally efficient method, we observe some model support $\log_{10}\Delta\mathcal{L}^{\rm HD}_{\rm CRN} = 1.1$ for the Hellings--Downs ORF, from which we derive the significance $p$-value above. If the signal is genuine, this result may indicate excess noise affecting the autocorrelations, or other weaknesses of the model.
\end{itemize} 

Some of these findings may be resolved by invoking anisotropy in the SMBHB source distribution, as simulations suggest a large variance is possible in the spectrum of a GWB \citep{Rosado+15}. The recovered amplitude of the common noise is close to the maximum value expected of a GWB from a population of SMBHBs \citep{Zhu+19}. The eccentricity of a nearby source or an unresolved local population may result in apparent the nonstationarity of the common noise. An analysis with a larger number of pulsars, for example under the IPTA, should be able to help answer these questions. It will also be interesting to investigate if other PTA observations of our \textit{quiet} pulsars (e.g. J1713+0747, J1744$-$1134) yield similar results under consistent noise models, and whether the temporal properties of the common noise and its upper limits are observed with independent sets of pulsars. Most importantly, it will be interesting to see if the global IPTA combination of the constituent PTA datasets strengthens the significance of any spatial correlations, resulting in an unambiguous detection. 

\begin{acknowledgments}

Murriyang, the Parkes 64\,m radio telescope is part of the Australia Telescope National Facility (\url{https://ror.org/05qajvd42}), which is funded by the Australian Government for operation as a National Facility managed by CSIRO. We acknowledge the Wiradjuri People as the Traditional Owners of the Observatory site. We acknowledge the Wurundjeri People of the Kulin Nation and the Wallumedegal People of the Darug Nation as the Traditional Owners of the land where this work was primarily carried out. We thank the Parkes Observatory staff for their support of the project for nearly two decades. We thank the International Pulsar Timing Array Detection Committee for their contributions to the IPTA GW search papers. We also thank CSIRO Information Management and Technology High Performance Computing group for access and support with the petrichor cluster. We acknowledge the use of the Python packages \textsc{numpy} \citep{numpy}, \textsc{scipy} \citep{scipy}, \textsc{chainconsumer} \citep{chainconsumer}, \textsc{corner} \citep{corner}, and \textsc{KDEpy} \citep{kdepy} for parts of this work. Part of this research was undertaken as part of the Australian Research Council (ARC) Centre of Excellence for Gravitational Wave Discovery (OzGrav) under grant CE170100004. R.M.S acknowledges support through ARC Future Fellowship FT190100155. Work at Naval Research Laboratory (NRL) is supported by NASA. SD is the recipient of an Australian Research Council Discovery Early Career Award (DE210101738) funded by the Australian Government. Y.L. acknowledges support of the Simons Investigator Grant 827103
\end{acknowledgments}

\end{document}